\documentclass[%
 aip,
 amsmath,amssymb,
reprint,%
]{revtex4-1}

\usepackage{graphicx}
\usepackage{dcolumn}
\usepackage{bm}

\usepackage[utf8]{inputenc}
\usepackage[T1]{fontenc}
\usepackage{mathptmx}

\usepackage{color}

\newcommand{\ie}{\textit{i.e.,}\ }
\newcommand{\eg}{\textit{e.g.,}\ }
\emergencystretch=\maxdimen
\begin{document}

\title{Modulation-Slippage Tradeoff in Resonant Four-Wave Upconversion}
\author{A. Griffith}
\email{arbg@princeton.edu}
\author{K. Qu}
\author{N. J. Fisch}
\affiliation{Department of Astrophysical Sciences, Princeton University,\\ Princeton, New Jersey 08540, USA}
\date{\today}

\begin{abstract}
	Following up on a proposal to use four-wave mixing in an underdense plasma at mildly relativistic laser intensities to produce vastly more energetic x-ray pulses [V.~M.~Malkin and N.~J.~Fisch, Phys. Rev. E, {\bf 101}, 023211 (2020)], we perform the first numerical simulations in one dimension to demonstrate amplification of a short high frequency seed through four-wave mixing.
	We find that parasitic processes including phase modulation and spatial pulse slippage limit the amplification efficiency.
	We numerically explore the previously proposed ``dual seed'' configuration as a countermeasure against phase modulation.
	We show how this approach tends to be thwarted by longitudinal slippage. 
	In the examples we considered, the best performance was in fact achieved through optimization of signal and pump parameters in a ``single seed'' configuration.
\end{abstract}

\maketitle
\section{Introduction}
Directly producing a megajoule of coherent x rays greatly exceeds capability of current x-ray technologies such as free electron lasers~\cite{emma_first_2010, pellegrini_physics_2016} or Compton scattering~\cite{powers_quasi-monoenergetic_2014}.
An alternative route for producing high-power coherent x rays is through efficient conversion of megajoule ultraviolet pulses, which are available at existing facilities like the National Ignition Facility~\cite{haynam_national_2007}.
However, conventional frequency conversion processes, such as high harmonic generation in gases~\cite{ferray_multiple-harmonic_1988, krausz_attosecond_2009, popmintchev_bright_2012}, crystals~\cite{ghimire_observation_2011, ghimire_high-harmonic_2019}, or relativistic plasma surfaces~\cite{RevModPhys.81.445}, cannot scale to the appropriate intensity or efficiency in the x-ray regime.

The challenges of achieving efficient frequency upconversion of high-power lasers can be overcome by working in plasmas.
Plasmas can resist the high intensities and high temperatures that disrupt solid or gaseous mediums.
Plasmas allow for wave-wave coupling processes, which have been investigated for laser amplification,
\eg Raman scattering~\cite{malkin_fast_1999-1,Ping2004,ren_compact_2008,pai_backward_2008,ping_development_2009,trines_simulations_2011,Vieux2011,Yampolsky2011,vieux_ultra-high_2017,PhysRevLett.120.024801,Balakin2020}, 
Brillioun Scattering~\cite{Lancia2010,lehmann_nonlinear_2013,riconda_spectral_2013-1,edwards_x-ray_2017,kirkwood_plasma-based_2018}, and magnetized scattering~\cite{Shi2018laser,edwards_laser_2019}.
Additionally, four-wave mixing in plasmas using atomic levels for frequency conversion~\cite{muendel_four-wave_1991} and pondermotive gratings for same frequency amplification~\cite{tang2019laser} have been considered.

In this paper, we  consider through numerical simulations the recent proposal to employ four-wave mixing in underdense plasma  to achieve both upconversion and amplification~\cite{malkin_towards_2020}.
In this proposal, a cascade of nonlinear, resonant four-wave interactions, based on a relativistic nonlinearity, was suggested to achieve up to a megajoule of laser energy in the x-ray regime~\cite{malkin_towards_2020}.
In each step of the cascade, two pump waves, at frequencies $\omega_1$ and $\omega_2$, amplify a weak higher frequency seed wave, frequency $\omega_3$.
An idler wave at frequency $\omega_4$ is generated to satisfy the resonance conditions,
\begin{align}
	&\omega_j^2 = \omega_{pe}^2 + c^2k_j^2,\qquad \omega_{pe}^2=\frac{4\pi n_ee^2}{m_e},\label{eqn:freq} \\ 
	&\omega_1 + \omega_2 = \omega_3 + \omega_4,\label{eqn:sync_time}\\ 
	&\mathbf{k_1} + \mathbf{k_2} = \mathbf{k_3} + \mathbf{k_4}.\label{eqn:sync_space}
\end{align}
Here,  the wave frequency $\omega_j$ corresponds to wavector $k_j$ ($j=1,2,3,4$) and plasma frequency $\omega_{pe}$, for an unperturbed electron fluid with particle charge $e$, mass $m_e$, and density $n_e$.
As the idler frequency may be small, the seed frequency for each iteration can reach up to the sum of the two pump frequencies.
Each interaction can thus give a multiplicative, rather than additive, change in frequency.
With iterated interactions, it might be possible to step up orders of magnitude in frequency.
It is the aim of this work to simulate one step of this cascade.

Ideally, the four-wave mixing can increase frequency with up to unity efficiency.
The maximum efficiency is achieved when the pumps are completely consumed.
If the symmetric pumps are ever depleted simultaneously, the four-wave interaction terminates. 
For synchronously depleted pumps all the wave energy is in the seed and the idler, and the pumps cannot regrow.
Thus the energy could, in principle, flow from the pumps to the seed and idler and never flow back to the pumps.
If the idler frequency is sufficiently low, it carries away negligible energy, resulting in almost all energy being consumed by the seed.
The unidirectional energy transfer possible in the four-wave mixing process~\cite{malkin_towards_2020} represents a significant advantage compared to three-wave scattering processes, which are susceptible to pump reamplification.

However, the elegant resonant four-wave interaction  becomes complicated when taking into account phase modulation.
Phase modulation changes the wave frequencies asynchronously with amplitude, thereby pushing the interaction out of resonance.
The same nonlinearity in the transverse direction can also result in filamentation of the pumps or seed.
To counteract the problems arising from phase modulation, the initial proposal suggested using a second signal and idler pair, namely a dual-seed approach, to balance the self- and cross-beam phase modulation terms against each other~\cite{malkin_towards_2020}.

In this paper, we confirm, using numerical simulations, that significant seed pulse amplification can occur through four-wave mixing, but the efficiency is limited by phase modulation.
The four-wave mixing process is additionally complicated by variable group velocities and dynamic envelope amplitudes.
The difference in group velocity tends to make the dual seed approach ineffective.
Instead, better performance is achieved, at least for the cases considered  here, with a single seed through selection of the four-wave parameters to minimize phase modulation.
We note that the cases that we consider are chosen, in part, for their ease in simulation, and do not represent the full range of possibilities.
However, the cases that we consider do expose both the upside potential and the key physical issues in realizing four-wave upconversion in underdense plasma.

\section{Resonance Conditions}

\begin{figure}
	\centering
	\includegraphics[width=\columnwidth]{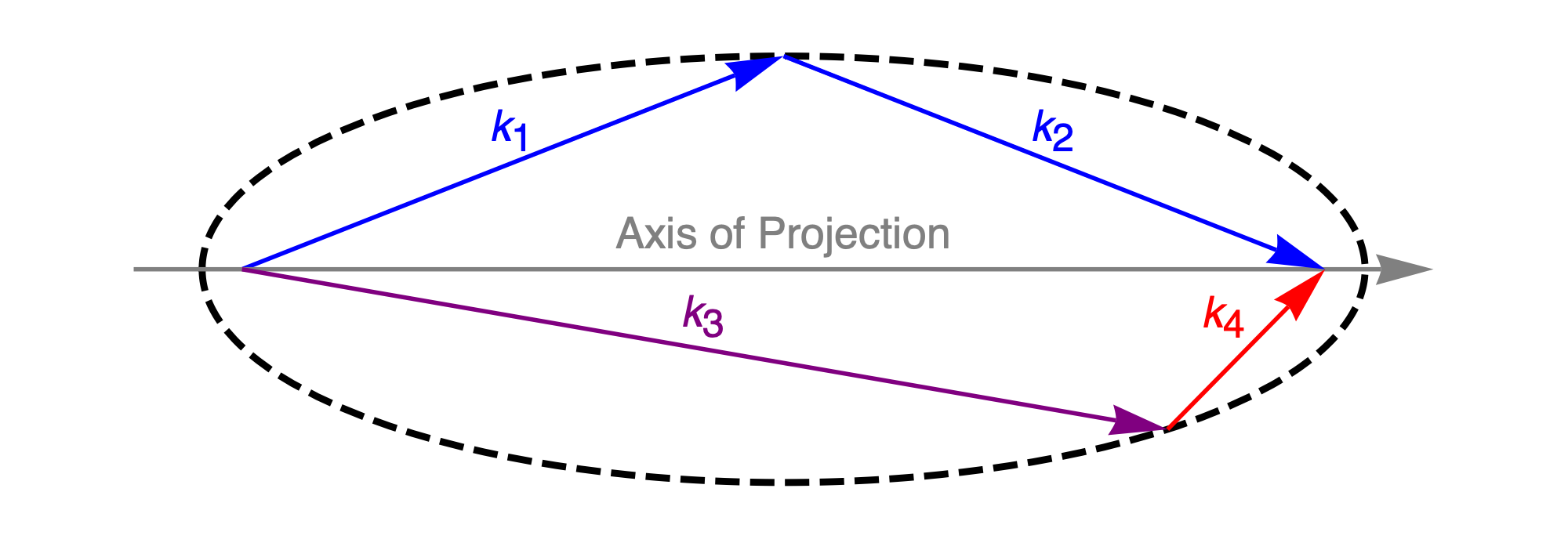}
	\caption{The dashed ellipse defines the resonance condition, setting the possible wave vector pairs, (1,2) and (3,4). 
	Simulations in one dimension are projected onto a single axis (gray) along the midline of the ellipse.}
	\label{fig:match}
\end{figure}

The resonance conditions, Eqns. (\ref{eqn:sync_time})-(\ref{eqn:sync_space}), determine the frequencies and propagation velocities of the four waves.
Four parallel waves are not desirable because they yield only two sets of trivial solutions: either $\omega_1=\omega_3$ corresponding to no frequency upconversion, or $\omega_{pe}= 0$ corresponding to no four-wave coupling.
Both solutions defeat the purpose of laser frequency upconversion. 
Satisfying Eqns. (\ref{eqn:sync_time}) and (\ref{eqn:sync_space}) while achieving frequency upconversion with a finite coupling coefficient requires misalignment.
Here, we note that frequency upconversion with colinear laser wavevectors is indeed possible when additionally manipulating the phase modulation\cite{malkin2020resonant}, but this approach is beyond the scope of our current paper.

The valid seed wavevectors, $\mathbf{k_3} = (k_{3_\parallel}, k_{3_\perp})$, form an ellipse determined by the pump wavectors, $\mathbf{k}_1$ and $\mathbf{k}_2$.
For convenience, we rotate the frame such that $k_{1_\perp} = -k_{2_\perp}$, and define quantities $2k = |\mathbf{k_1}+\mathbf{k_2}| = k_{1_\parallel}+k_{2_\parallel}$ and $2\omega = \omega_1 + \omega_2$,
\begin{equation}
	1 - \frac{\omega_{pe}^2}{\omega^2-c^2k^2} = \frac{c^2(k_{3_\parallel}-k)^2}{\omega^2}+ \frac{c^2k_{3_\perp}^2}{\omega^2-c^2k^2}.
	\label{eqn:cond}
\end{equation}
The ellipse, as illustrated in Fig.~\ref{fig:match}, represents the complete set of pump-pump and signal-idler wavevector pairs.
It has two foci, located at
\begin{equation}
    k \left(1\pm\sqrt{1-\frac{\omega_{pe}^2}{\omega^2-c^2k^2}}\right).
\end{equation}
The seed frequency is maximized when $\mathbf{k_3}$ extends beyond the right focus and touches the rightmost point on the ellipse, \ie
\begin{equation}
    \max|\mathbf{k}_3| = k+c^{-1}\omega\sqrt{1 - \frac{\omega_{pe}^2}{\omega^2-c^2k^2}}.
    \label{eqn:max_w3}
\end{equation}
For fixed pump frequencies, the maximum seed frequency is achieved when the pumps are misaligned by an angle
\begin{equation}
    \theta_{1,2} \approx \sqrt{(c|\mathbf{k_1}| |\mathbf{k_2}|)^{-1}(|\mathbf{k_1}| + |\mathbf{k_2}|)\omega_{pe}}. 
    \label{eqn:theta}
\end{equation}
Note that, in depicting the ellipse  in Fig.~\ref{fig:match}, in contrast to the limiting case portrayed in previous work~\cite{malkin_towards_2020}, all wave vectors are not necessarily chords on the ellipse.
The wave vectors only approach chords in the high frequency paraxial limit when both $\omega_{pe}^2/(\omega^2-c^2k^2)$ and $1-\omega^{-2}c^2k^2$ vanish simultaneously.
In this limit, the ellipse becomes long and thin, and approaches a line segment between the foci at $0$ and $2k$.
Depending on the magnitude of the plasma frequency and the angle between the pumps, the origin and end point of the $\mathbf{k_1}$ and $\mathbf{k_2}$ or $\mathbf{k_3}$ and $\mathbf{k_4}$ pairs may lie inside or outside the ellipse.

The required wavevector misalignment results in slippage between the four waves.
The angles, and consequently velocities, between the rest of the wavevectors are best interpreted through considering Fig.~\ref{fig:match}.
As $\omega_3$ increases, it pulls the tip of $\mathbf{k_3}$ towards the rightmost point of the ellipse, becoming more parallel with the major axis.
To satisfy the resonance conditions, $\mathbf{k_4}$ must correspondingly tilt more inward.
There is a resulting ordering of $|k_{4_\perp}/k_{4_\parallel}|>|k_{1,2_\perp}/k_{1,2_\parallel}|>|k_{3_\perp}/k_{3_\parallel}|$.
In the projection, the misalignment drives $v_{3_\parallel} > v_{1,2_\parallel} > v_{4_\parallel}$, causing a slippage between the waves.

The slippage can be reduced with smaller pump laser angles, but plasma dispersion must increase to keep the four-wave coupling rate constant.
The four-wave coupling rate~\cite{malkin_towards_2020} scales with both the angle between the pump beams and the plasma frequency, \ie
\begin{equation}
    	 \frac{k_{1,2_\perp}^2\omega_{pe}^2}{k_{1,2}^2\sqrt{\omega_{3}\omega_4}} \left|\frac{eA_{1,2}}{m_ec^2}\right|^2.
    	 \label{eqn:coupling}
\end{equation}
Decreasing misalignment and dispersion both reduce the four-wave coupling through $k_{1,2_\perp}/k_{1,2}$ and $\omega_{pe}^2/\sqrt{\omega_3\omega_4}$ respectively.
A decrease in either term may be compensated for through increasing the magnitude of the pump vector potential, $A_{1,2}$.
But the compensation is capped as pump strength may only grow as long as $|eA_{1,2}| \ll m_ec^2$ to remain in the mildly relativistic regime.
The perpendicular wavevector component and dispersion contribute similarly to the parallel velocity,
\begin{equation}
    v_{j_\parallel}/c \approx 1 -k_{j_\perp}^2k_j^{-2}- \omega_{pe}^2(2c^2k_j^2)^{-1}.
\end{equation}
Either misalignment, $k_{j_\perp}^2k_j^{-2}$, or dispersion, $\omega_{pe}^2(2c^2k_j^2)^{-1}$, may be small, but not both if strong coupling is desired.

The misalignment is the dominant effect for large upconversion.
For large upconversion, $k_{1,2_\perp} \approx k_{1,2}\theta_{1,2}/2$.
With $\theta_{1,2}$ chosen  in accordance with Eqn.~(\ref{eqn:theta}), the missalignment term contributes slippage linear in $\omega_{pe}/ck_{1,2}$.
The missalignment slippage term which is linear in $\omega_{pe}$ dominates the dispersion term which is quadratic in $\omega_{pe}$ as the waves remain in the underdense regime.
The slippage due to differences in misalignment can be demonstrated to have a significant effect on the four-wave upconversion process.


\section{Four-Wave Model}
The four-wave interaction is governed by a set of nonlinear wave equations derived through combining Maxwell's equations, the relativistic equations of motion for a constant density mono-energetic electron fluid, and the neutralizing effect of a static ion background.
Each wave has a scaled complex envelope $b_j = \sqrt{\omega_j} eA_j/(m_ec^2)$, where $A_j$ is the vector potential amplitude for wave $j$. 
All the waves are polarized perpendicular to the plane in which all $\mathbf{k_j}$ are chosen to lie.
The four-wave interaction originates from the lowest order relativistic correction to the electron equations of motion, expanding in $ eA_j/(m_ec^2)$.

To pose the problem in 1D (one dimension), the dynamical equations\cite{malkin_towards_2020} are projected onto the dominant axis of propagation.
The axis, denoted $x$, lies on the center of the ellipse governed by Eqn.~(\ref{eqn:cond}), chosen such that $k_{\perp_j}/k_{\parallel_j} \ll 1$ for all waves.
The resulting dynamical equations contain longitudinal propagation and the lowest order relativistic correction,
\begin{align}
	  & i(\partial_t + c^2 k_{j_x}\omega_j^{-1}\partial_x)b_j = \delta\omega_j b_j + \partial_{b^*_j}\mathcal{H},
	\label{eqn:evol}\\
	  & \delta\omega_j = \frac{\omega_{pe}^2}{2\omega_j}\sum_{l=1}^{4}
	\frac{|b_l|^2}{\omega_l}\times\begin{cases}
		f_{+, j, l} + f_{-,j,l} - 1, & j \neq l \\
		f_{+, j, l} - 1,             & j = l
	\end{cases} \label{eqn:mod}\\
	  & \mathcal{H} = V_{1,2,3,4}b_1b_2b_3^*b_4^* + c.c.,\\
	  & V_{j,l,m,n} = \frac{\omega_{pe}^2(f_{+,j,l}+f_{-,j,n}+f_{-,l,n})}{2\sqrt{\omega_j\omega_k\omega_m\omega_n}},   \\
	  & f_{\pm, j, l} = \frac{c^2(\mathbf{k_j}\pm\mathbf{k_l})^2}{(\omega_j\pm\omega_l)^2-\omega_{pe}^2}- 1. \label{eqn:fjl}
\end{align}
The L.H.S. of Eq.~(\ref{eqn:evol}) describes the wave propgation in the $x$ direction at group velocity $v_j=c^2k_{j_\parallel}\omega_j^{-1}$.
The R.H.S. consists of a modulational term, $\delta \omega_j$, which results in amplitude dependent frequency shifts, and a four-wave coupling term, captured through the Hamiltonian $\mathcal{H}$.

The paraxial equations~(\ref{eqn:evol}) are evolved numerically to capture the long time amplification of the seed.
We conduct the simulations in a frame moving with the seed to reduce the computational domain.
The two pump waves are initialized evenly out in front of the signal seed, and the seed runs through the waves picking up their energy.
The fourth idler wave, which has the smallest parallel group velocity, quickly flows out of the left side of the domain.
The equations are evolved using Dedalus, a general spectral PDE solver ~\cite{burns_dedalus_2020}.

\begin{figure}
	\centering
	\includegraphics[width=\columnwidth]{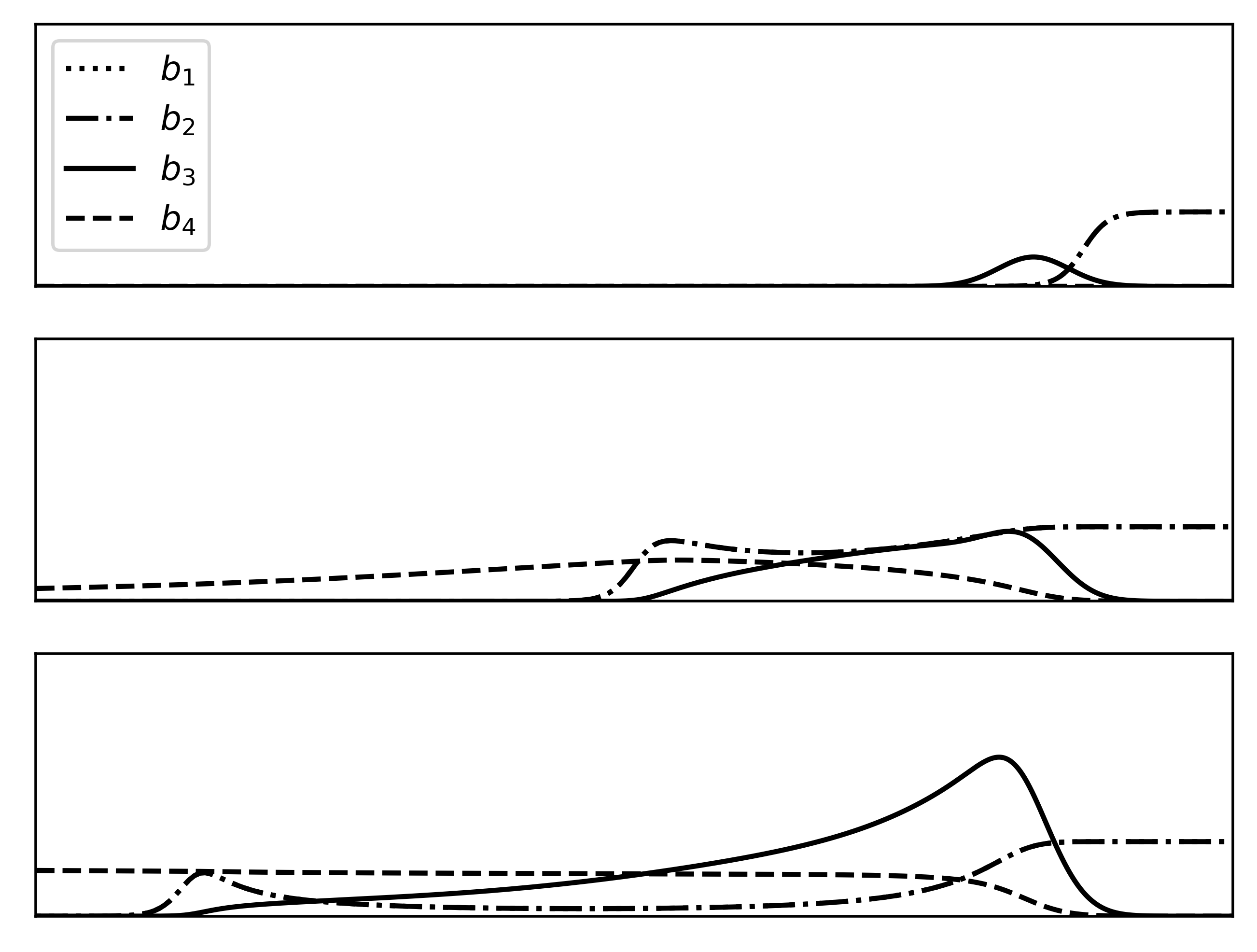}
	\caption{Numerical snapshots of $b_j$ with upconversion factor $\omega_3/\omega_{1,2} = 1.4$.
	Snapshots are shown at $\omega_{pe}t = 0,\ 10^4,\ 2\times 10^4$.
	Pumps ($b_1$, dotted, and $b_2$, dot-dashed) are fed in with relativistic factor $|a_{1,2}| = 0.33$, with the seed initialized with relativistic factor $|a_3| = 0.1$.} 
	\label{fig:amp_no_mod}
\end{figure}

\section{Ideal Four-Wave Behavior}
To illustrate the opportunities in four-wave resonant mixing, we first simulate an ideal scenario for the four-wave interaction. 
In this simulation, phase modulation is assumed to be negligible, which will expose the successes and challenges intrinsic to  four-wave resonant coupling.
Consider then pumps that have the same frequency, but with equal and opposite $k_\perp$.
The resulting synchronous pumps amplify the seed, which grows monotonically in energy. 

Figure~\ref{fig:amp_no_mod} shows three snapshots of a seed being amplified by several orders of magnitude in total energy.
The first snapshot shows the initial conditions (held constant across all simulations).
The pumps are just beginning to intersect with the seed and initiate the linear stage of amplification.
When the pumps are still strong, amplification occurs widely, resulting in the long seed tail shown in the second snapshot.
As the seed strength grows the pumps become depleted, and amplification occurs closer to the front of the seed.
For a sufficiently strong seed, all of the pump energy is consumed at the seed's leading edge.
The signal then grows continually steeper in time, taking long duration pump energy and compressing it into a shorter peak.
The compression of pump energy is similar to that in Raman amplification \cite{malkin_fast_1999-1}.
Like Raman amplification, some energy is lost to a disposable wave, the fourth wave here, or the plasma wave in the case of  Raman amplification.
But, unlike Raman amplification, all energy could ideally be deposited into a single growing peak, without producing the amplified pulse train characteristic of the $\pi$ pulse solution for resonant 3-wave interactions ~\cite{malkin_fast_1999-1}.

Thus, considerable upshift and efficiency are easily achieved in the idealized four-wave resonant interaction. 
The results shown in Fig.~\ref{fig:amp_no_mod} achieve a $40\%$ increase in pump photon energy with $\omega_1+\omega_2 \sim 10^2 \omega_{pe}$.
When pump depletion is achieved, the energy conversion efficiency may become as high as $70\%$, with the remaining energy flowing to the idler wave. 
The frequency-upshifted output wave is advantageously single-peaked.

\begin{figure}
	\centering
	\includegraphics[width=\columnwidth]{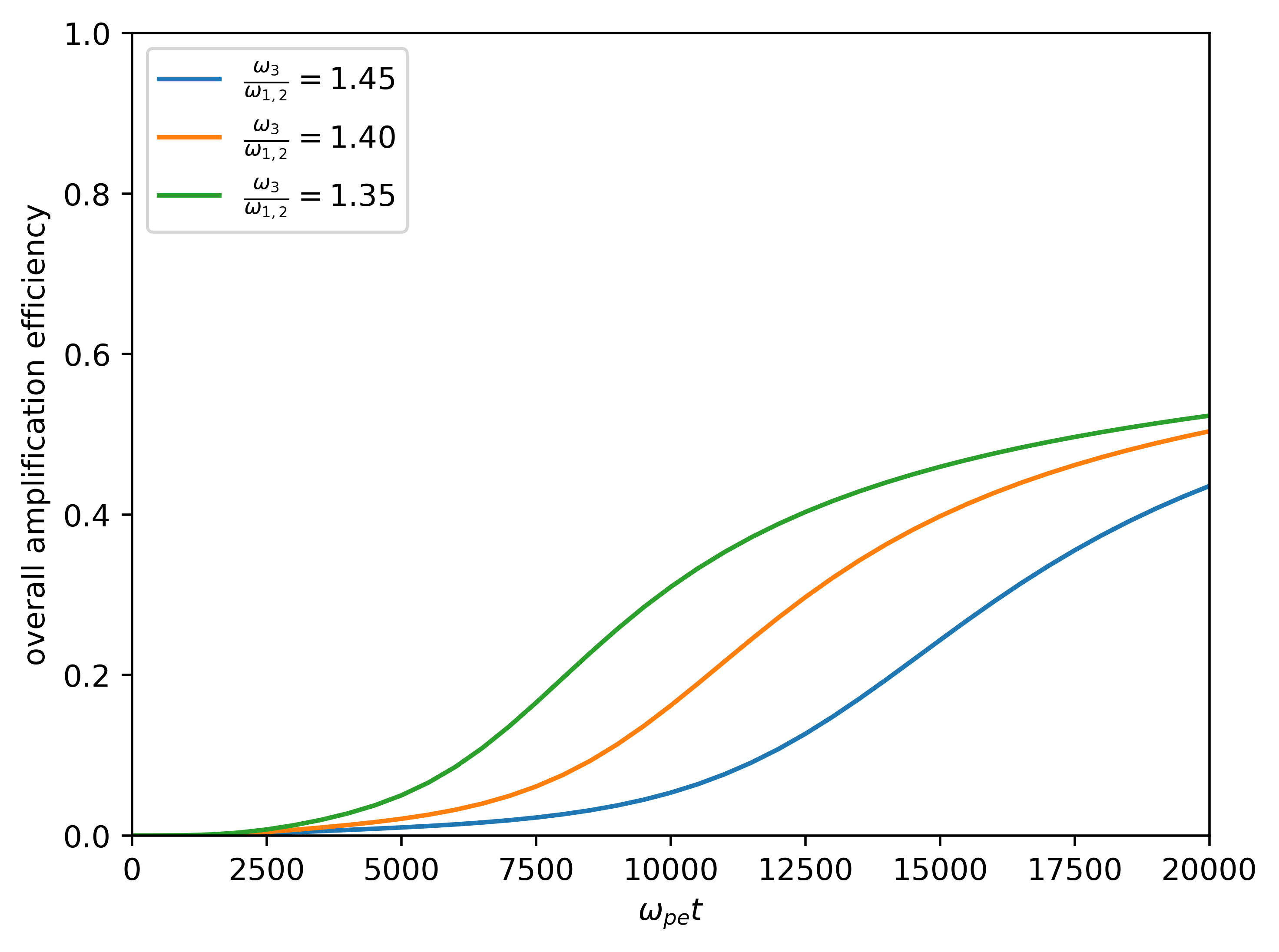}
	\caption{Increasing $\omega_3/\omega_{1,2}$ decreases coupling, resulting in a longer time to efficiency saturation.
	Over a finite timescale this reduces efficiency, even though higher saturation efficiency may be reached.
	$\omega_{3} = 1.4\omega_{1,2}$ corresponds with evolution shown in Fig. \ref{fig:amp_no_mod}. 
	All efficiencies are simulation with Fig. \ref{fig:amp_no_mod} parameters, varying $\omega_3$. $\omega_3 = 1.35, 1.40, 1.45\omega_{1,2}$ correspond to ideal efficiencies of $67.5, 70.0,72.5 \%$ respectively.}
	\label{fig:eff_freq}
\end{figure}

A larger seed frequency corresponds to a higher limiting efficiency, but, in practice,  this higher efficiency is difficult to achieve within a finite plasma length.
At higher seed frequency, upconversion is hampered by the consequent decrease in the coupling coefficient.
Figure~\ref{fig:eff_freq} shows the reduction in the realized amplification efficiency with increasing seed frequency.
Weaker coupling increases the time to reach the pump depletion regime.
Only at pump depletion is maximum efficiency and steepening achieved.
For a limited interaction region, the maximum efficiency may never be reached, and upconversion may be strictly worse for higher frequencies.

\section{Phase Modulation and the Four-Wave Interaction}
The ideal solution, however, neglected phase modulation.  
The phase modulation terms must in fact be included to capture  fully the lowest order relativistic behavior.    
These terms can push the four-wave interaction out of resonance.
For example, the issue caused by phase modulation can be seen in the case of the exact same wavevectors used in the ideal regime, e.g. Fig.~\ref{fig:amp_no_mod}.   
The same frequency pumps cancel in the denominator in Eqn. (\ref{eqn:fjl}), making phase modulation scale $\propto c^2k_{1_\perp}^2\omega_{pe}^{-2}$.
A large value of $ck_{1_\perp}\omega_{pe}^{-1}$ is required for significant four-wave coupling, resulting in extreme phase modulation.
For the parameters used in Fig. \ref{fig:amp_no_mod}, the strength of this term results in a perturbation from resonance that makes amplification untenable, with 
$\delta\omega_{1,2}$ approximately 40 times the seed growth rate.
The wavevectors must then change for the interaction to coherently amplify the seed.

\begin{figure}
	\centering
	\includegraphics[width=\columnwidth]{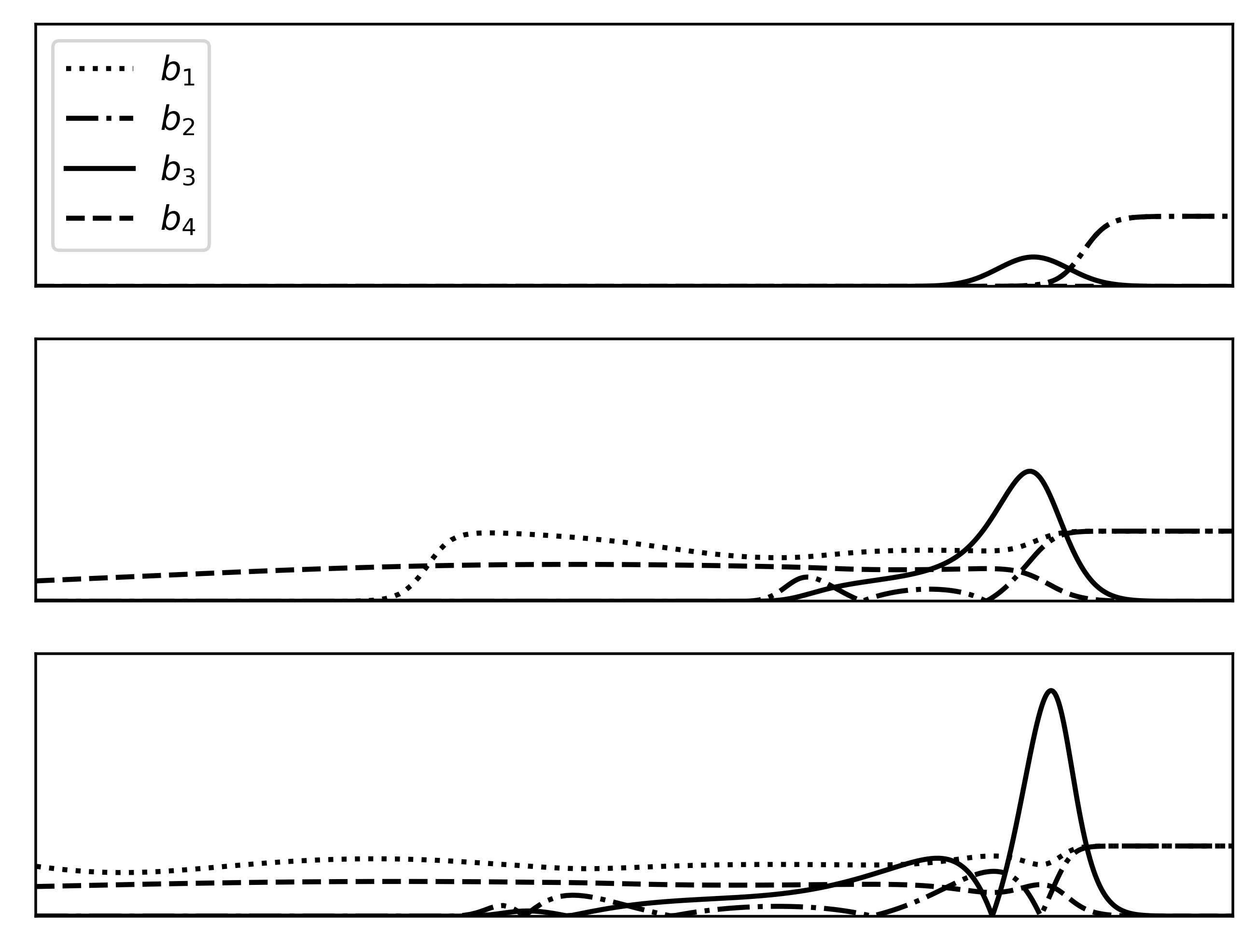}
	\caption{Numerical snapshots of $b_j$ at $\omega_{pe}t = 0,\ 10^4,\ 2\times 10^4$ for $2\omega_3/(\omega_{1}+\omega_2) = 1.4$ with $\delta\omega_j$ set to $0$, but with pumps detuned such that$\omega_2 - \omega_1 = 11\omega_{pe}$. Pumps are initialized with $|a_{1,2}| = 0.33$ and the seed is given amplitude $|a_3| = 0.1$.}
	\label{fig:4wave_no_mod}
\end{figure}

\begin{figure}
	\centering
	\includegraphics[width=\columnwidth]{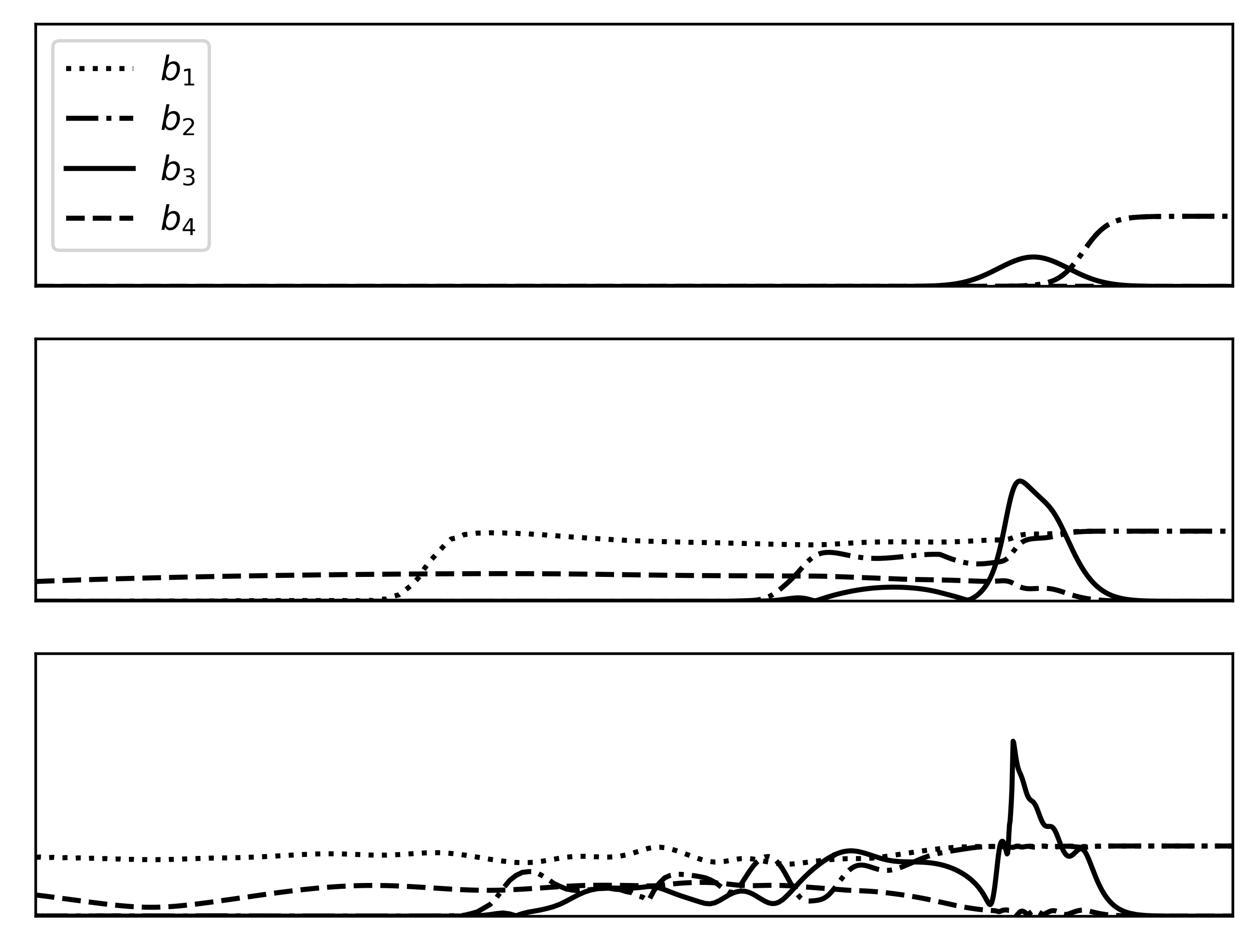}
	\caption{Numerical snapshots of $b_j$ at $\omega_{pe}t = 0,\ 10^4,\ 2\times 10^4$ for $2\omega_3/(\omega_{1}+\omega_2) = 1.4$ with $\delta\omega_j$ governed by Eqn. (11), but with pumps detuned such that$\omega_2 - \omega_1 = 11\omega_{pe}$. Pumps are initialized with $|a_{1,2}| = 0.33$ and the seed is given amplitude $|a_3| = 0.1$.}
	\label{fig:4wave_mod}
\end{figure}

The resonance drift caused by phase modulation can be reduced through pump detuning.
Pump detuning reduces the strength of the $f_{-, 1, 2}$ term to $\propto c^2k_{1_\perp}^2(\omega_1-\omega_2)^{-2} \ll  c^2k_{1_\perp}^2\omega_{pe}^{-2}$ for $\omega_1 - \omega_2 \gg \omega_{pe}$.
But detuning the pumps, while necessary to reduce modulation, has a corresponding cost in pump-pump slippage.
The frequency detuned pumps have different velocities relative to the seed,  so that they no longer move in perfect unison.
As a result, the perfectly simultaneous pump depletion of the ideal case is no longer possible.
To isolate the indirect effects, namely slippage, from the direct effects, namely phase modulation, we performed simulations with the same detuned pumps, both excluding and including the $\delta\omega$ term.
The results illustrated by Fig.~\ref{fig:4wave_no_mod} show that asynchronous pumps can cause the four-wave interaction to work in reverse, leading to re-amplified pumps and reduced energy conversion efficiency.
In Fig. \ref{fig:4wave_mod} phase modulation is added into the same simulation.
The phase modulation becomes significant at high seed amplitude, pushing the waves out of resonance, and further lowering amplification.

\begin{figure}
	\centering
	\includegraphics[width=\columnwidth]{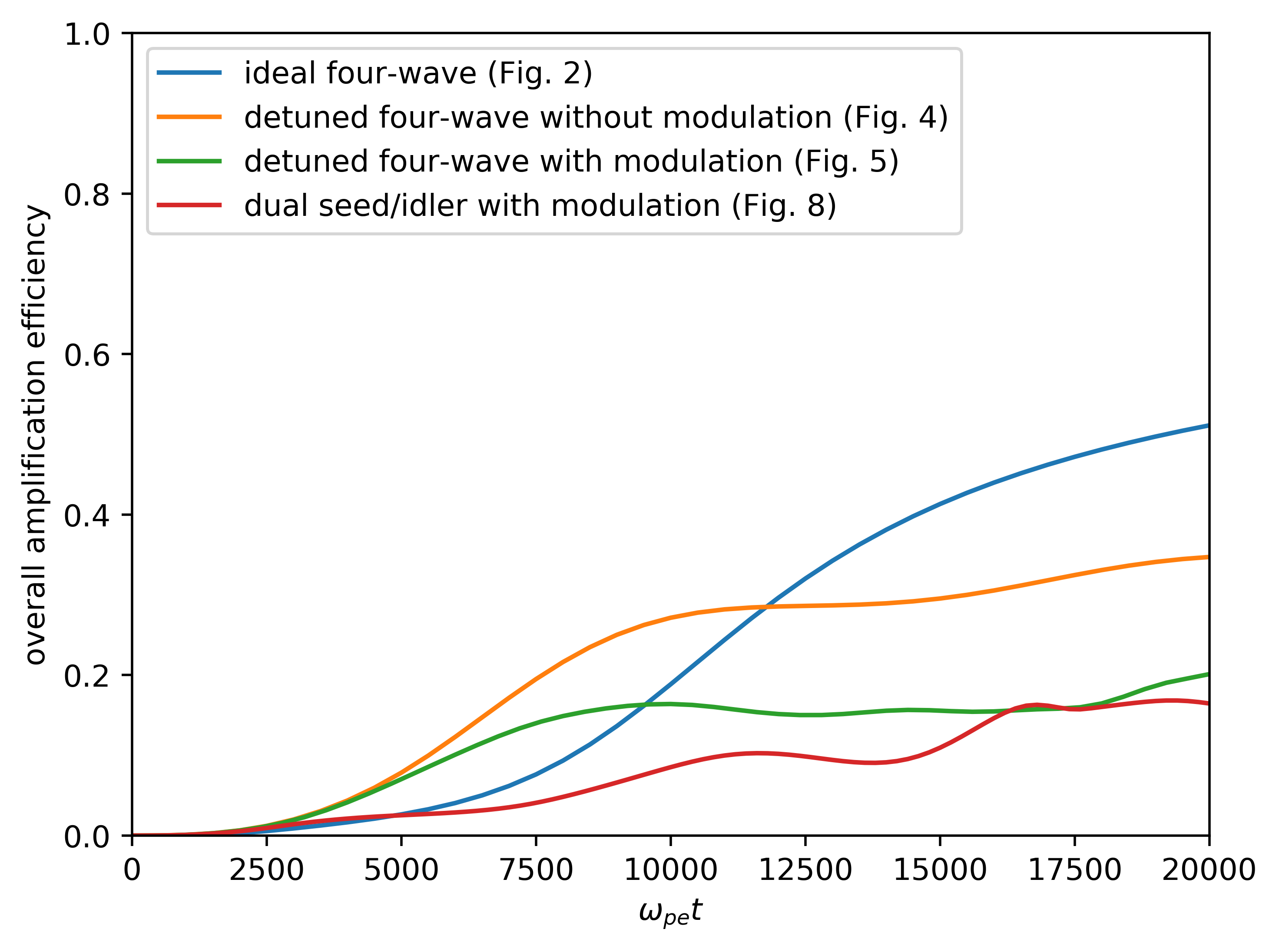}
	\caption{Efficiency evolution corresponding to snapshots presented in Figs. \ref{fig:amp_no_mod}, \ref{fig:4wave_no_mod}, \ref{fig:4wave_mod}, and \ref{fig:six_wave}.
	Symmetric pumps equilibrate at a much longer timescale, but at higher level compared to non-ideal alternatives.
	All cases correspond with $2\omega_3/(\omega_1+\omega_2) = 1.4$.}
	\label{fig:mod_comp_eff}
\end{figure}

The combination of detuning and phase modulation apparently sets a lower achievable maximum efficiency.
The efficiency evolution of the detuned simulations with and without phase modulation both perform worse than the earlier ideal simulations, shown in Fig. \ref{fig:mod_comp_eff}.
The efficiency rises faster in the detuned simulations, as detuning moderately increases coupling.
However, it becomes bounded at a lower level from detuning, and is driven even lower from phase modulation.
Both factors contribute significantly, with slippage on its own driving a large change in efficiency.
The importance of slippage as an indirect effect will persist even as we attempt to mitigate the modulation through other means.

\section{Counterbalancing Phase Modulation with dual seeds}
Following the original suggestion~\cite{malkin_towards_2020}, to reduce the detrimental effects of modulation, the four-wave approach can be qualitatively changed through adding two more waves. 
The additional beams induce cross-beam modulation which could, in principle, counterbalance against the pump/pump phase modulation.
Cross-beam phase modulation between slightly detuned seeds can oppose the cross-beam phase modulation between the slightly detuned pumps.

The two additional waves obey the same resonance conditions (Eqns. (\ref{eqn:ext_sync_time}) and (\ref{eqn:ext_sync_space})), where wave $5$ will be the second seeded signal and a wave $6$ will be the second idler.
\begin{align}
	  & \omega_1 + \omega_2 = \omega_3 + \omega_4 = \omega_5 + \omega_6,\label{eqn:ext_sync_time}                          \\
	  & \mathbf{k_1} + \mathbf{k_2} = \mathbf{k_3} + \mathbf{k_4} = \mathbf{k_5} + \mathbf{k_6}.\label{eqn:ext_sync_space}
\end{align}
The dynamical equations must be adjusted to account for the two new waves. 
Evolving two additional waves adds additional phase modulation terms,
\begin{align}
	 &\delta\omega_j = \frac{\omega_{pe}^2}{2\omega_j}\sum_{l=1}^{6}
	\frac{|b_l|^2}{\omega_l}\times\begin{cases}
		f_{+, j, l} + f_{-,j,l} - 1 & l \neq j \\
		f_{+, j, l} - 1             & l = j
	\end{cases} \label{eqn:six_mod}.
\end{align}
The phase modulation has been extended to all six waves, with the novelty primarily contained in the strength of the new $f_{-,3,5}$ term.
The similarly large $f_{-,4,6}$ doesn't significantly contribute as waves $4$ and $6$ never grow large, slipping behind the point of interaction much faster than signal and pump waves. 

The Hamiltonian governing the four-wave coupling must also be extended to accommodate waves $5$ and $6$.
\begin{align}
	 &\mathcal{H} = V_{1,2,3,4}b_1b_2b_3^*b_4^* + V_{1,2,5,6}b_1b_2b_5^*b_6^*\nonumber\\
	 &\qquad + V_{3,4,5,6}b_3b_4b_5^*b_6^* + c.c.
	\label{eqn:six_coupl}
\end{align}
A second set of four-wave coupling results in symmetric pump-pump to signal-idler transfer for waves five and six, $V_{1,2,5,6}$, as previously only was for waves three and four, $V_{1,2,3,4}$.
The resonance conditions imply a novel term which is the signal-idler to dual signal-dual idler coupling, $V_{3,4,5,6}$.
When both the $3,5$ and $4,6$ pairs are perfectly symmetric the additional coupling shouldn't change the four-wave behavior, but, when the signal waves begin to slip, the coupling can push energy between the desynchronized seeds.

\begin{figure}
	\centering
	\includegraphics[width=\columnwidth]{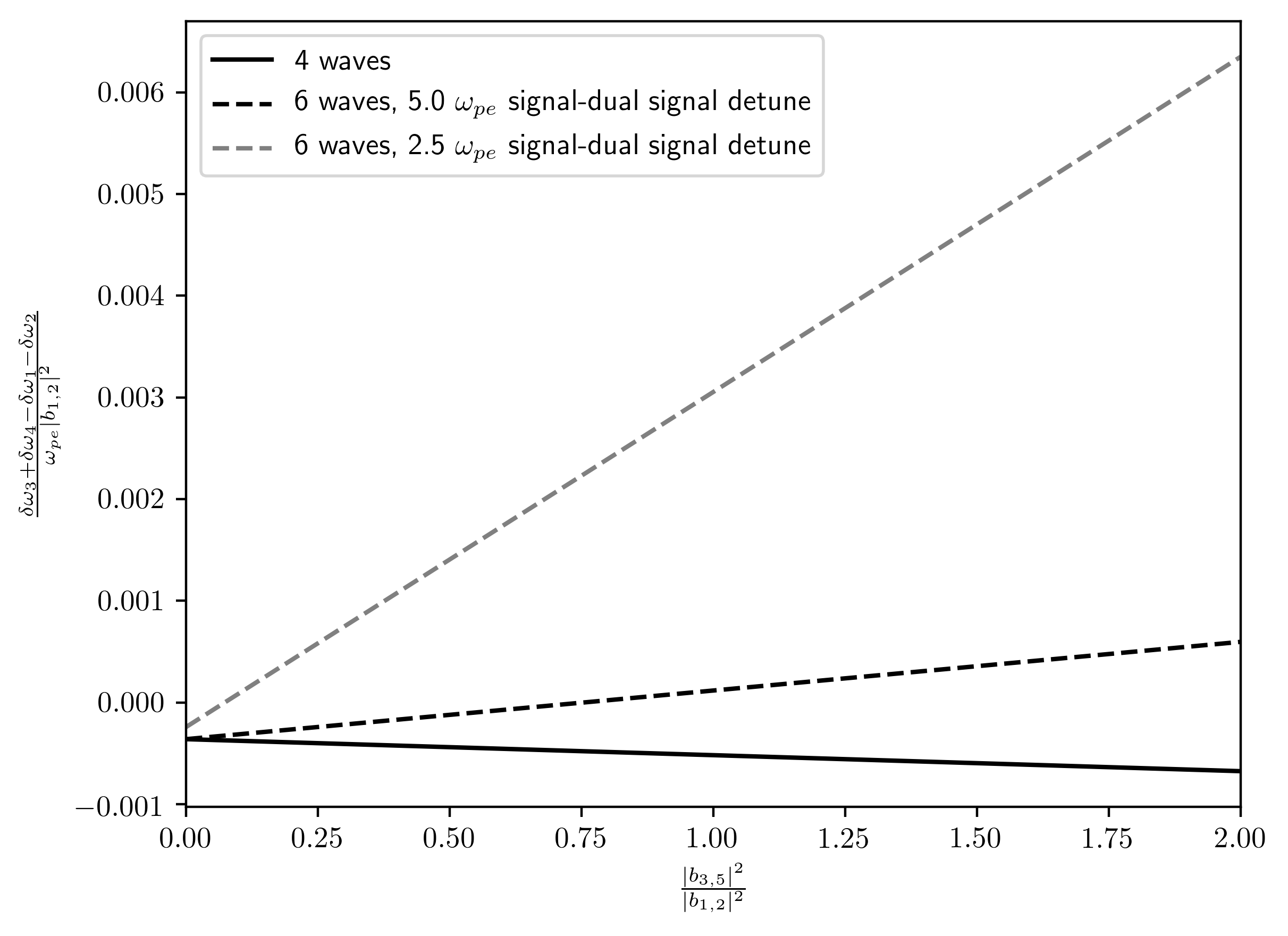}
	\caption{Resonance detuning, $\delta\omega_1 + \delta \omega_2 - \delta\omega_3-\delta\omega_4$, may have a different sign dependence on seed strength if the dual seed approach is used.
	A point of perfect resonance is achieved at finite signal and pump amplitude with appropriate application of the dual seed strategy.}
	\label{fig:mod_scenarios}
\end{figure}

The new phase modulation terms may be arranged, with specific finite wave amplitudes, such that the four waves are in perfect resonance.
The arrangement is accomplished through changing the detuning between the two seeds.
Changing the detuning alters the seed to pump ratio at which all phase modulation terms balance.
Two cases with $\omega_3-\omega_5 = 2.5\omega_{pe}, \text{ and } 5\omega_{pe}$ are compared to the unaltered scheme in Fig.~\ref{fig:mod_scenarios}.
Weak detuning results in extreme sensitivity of the resonance to the seed to pump ratio and a low relative seed strength at which the terms balance.
Larger detuning results in lower sensitivity, and counterbalancing at larger seed amplitude, where the counterbalancing is needed most.
Of course, the perfect resonance may be lost as the waves evolve in time away from the arranged amplitudes.

\begin{figure}
	\centering
	\includegraphics[width=\columnwidth]{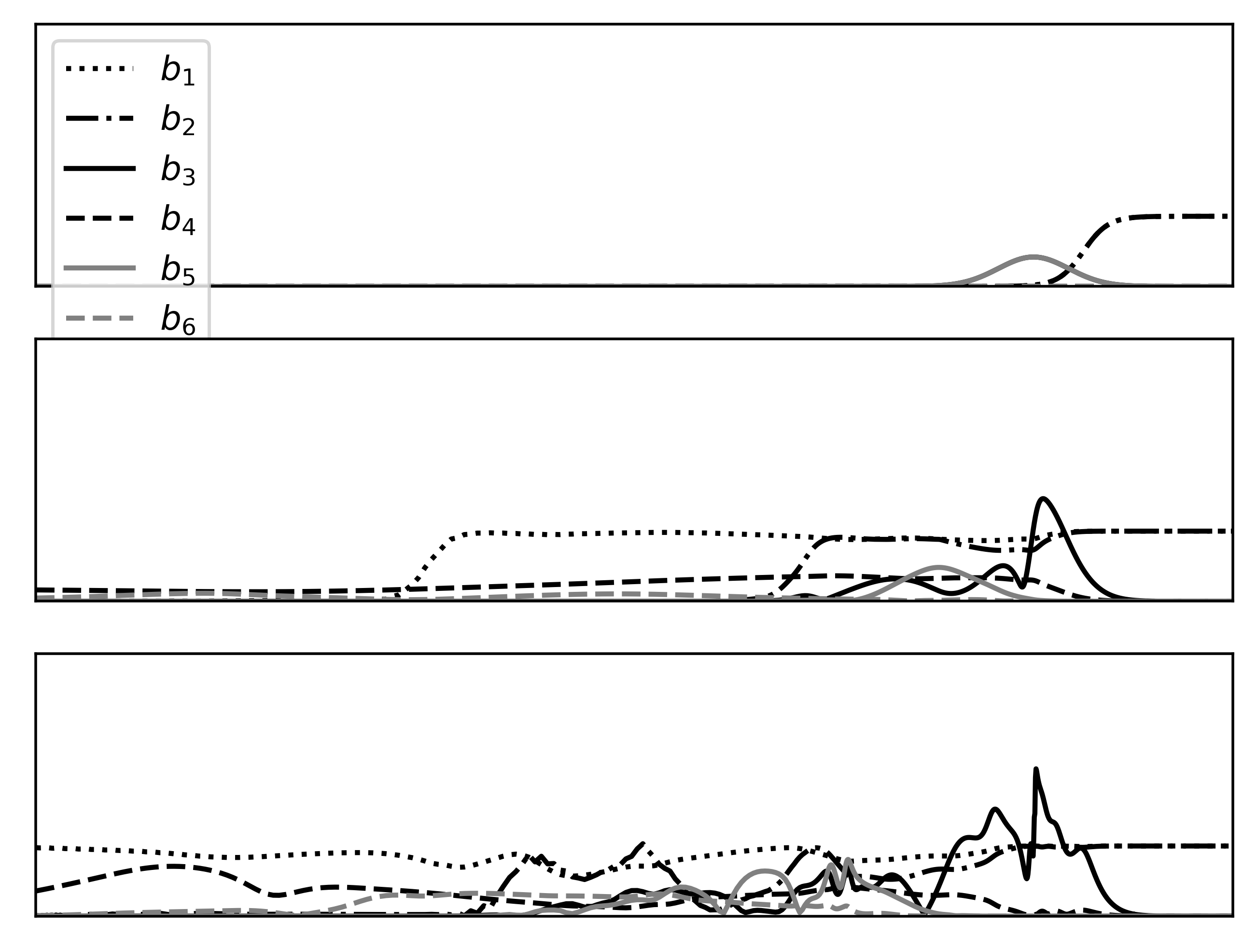}
	\caption{Numerical snapshots of $b_j$ at $\omega_{pe}t = 0,\ 10^4,\ 2\times 10^4$ for $2\omega_3/(\omega_{1}+\omega_2) = 1.4$ with $\omega_2 - \omega_1 = 11\omega_{pe}$ and with a second seed and idler such that $\omega_3 - \omega_5 = 5\omega_{pe}$. Pumps are initialized with $|a_{1,2}| = 0.33$ and the seed is given amplitude $|a_3| = 0.1$.}
	\label{fig:six_wave}
\end{figure}

Thus, a fifth and sixth wave are added to the previous simulations to evaluate the dual signal/idler approach. 
The new initial conditions are shown in the first snapshot of Fig.~\ref{fig:six_wave}.
Now, the second signal seed is given the same initial envelope as the initial signal, such that initially waves three and five completely overlap.
Wave five is detuned from wave three by $5\omega_{pe}$, and slips behind the leading seed as can be seen in the second snapshot.
Finite detuning results in group velocity differences, and the new lower frequency second signal wave falls behind on a faster timescale than the amplification.
The leading signal wave then gains more energy than the second signal.
The signal-to-signal coupling further amplifies this issue as it drives an energy transfer between the two signal waves.
The required symmetry between the two signal waves quickly fades, and the simulation begins to converge toward the earlier unaltered simulation, where the last snapshots of Fig.~\ref{fig:six_wave} and Fig.~\ref{fig:4wave_mod} have similar signal wave envelopes.

As with detuning the pumps, there is apparently an unavoidable tradeoff between reducing phase modulation and slippage.
The phase modulation dominantly affects the resonance through the seed-seed coupling, $f_{-,3,5}$.
The sensitivity of the resonance to the seed strength scales inversely with the detuning,
\begin{equation}
    \frac{\partial(\delta\omega_1 + \delta\omega_2 - \delta\omega_3-\delta\omega_4)}{\partial(|b_{3,5}|^2)} \sim \frac{2c^2k_{3_\perp}^2\omega_{pe}^2}{\omega_3^2(\omega_3-\omega_5)^2}.
\end{equation}
While the sensitivity of the resonance decreases with larger seed-seed detuning, the slippage between the two waves increases linearly,
\begin{equation}
    v_{3_\parallel}-v_{5_\parallel}\sim \frac{2c^3(\omega_3-\omega_5)k_{3_\perp}^2}{\omega_3^3}.
\end{equation}
Both low sensitivity and low slippage cannot be achieved simultaneously, but both are needed to make the six-wave approach effective.

The six-wave strategy in our implementation performs worse than the approach without the dual seed.
The worse performance is seen not only in pulse structure, in Fig.~\ref{fig:six_wave}, but also in the efficiency, as shown in Fig.~\ref{fig:mod_comp_eff}.
Only after the second seed has fully fallen behind the original seed does the efficiency begin to approach that of the unaltered four-wave approach.
In our implementation we have not found a parameter regime in which the six-wave solution improves upon the comparable simpler four-wave approach.

\section{Conclusions}
In conclusion, we described, using one-dimensional simulations, how idealized four-wave mixing with two balanced pumps, can amplify, with high efficiency, significantly upshifted pulses.
However,  this idealized case neglected phase modulation terms, which are difficult to cancel out.
%
%
When phase modulation is considered, the pump wavevectors must be changed to reduce cross-beam phase modulation.
With just a simple strategy to mitigate phase modulation, successful upconversion can be achieved, but with efficiency significantly lower than in the ideal scenario.
At least in the cases that we considered, adding a second signal-idler pair to mitigate phase modulation has not managed to achieve the theoretically available efficiencies.
The second seed's slippage results in asymmetry between seeds, which is further exacerbated by additional coupling.
The asymmetry between the seeds serves more to reduce the amplification  than to remove the limiting effects of phase modulation.
In all the cases that we considered, optimizing for low slippage was as important as optimizing for favorable four-wave coupling and phase modulation mitigation.

The considerations here are just a first cut at describing numerically the issues in optimizing the recently proposed four-wave coupling in underdense plasma to produce frequency upshifted laser power with very high efficiency.  
While the efficiencies reached in these simulations were considerable, they fell short of  the theoretically achievable  efficiencies.
Nonetheless, the possibilities explored here do not exhaust what might be attempted to achieve those theoretically achievable  efficiencies.
Other possibilities include introducing more waves to control the phase modulation and to vary the waveforms in space. 
In particular, we have not considered the suggestion of utilizing grazing angle reflections in a channel, which carries perhaps the major opportunities for realizing the potential of four-wave interactions~\cite{malkin_towards_2020}.  
While enlarging the parameter space of the most promising avenues to be explored, these possibilities do come, however, with added complexity in experimental realization and computational cost in simulations. 
What we did explore already showed both the very promising potential of the four-wave upconversion effect and the issues in realizing it.

\begin{acknowledgements}
The work is supported by NNSA Grant No. DE-NA0003871.

\section*{Data Availability}
The data that support the findings of this study are available from the corresponding author upon reasonable request.

\end{acknowledgements}

\bibliography{numerical_sim_4wave_paper}

\begin{thebibliography}{33}%
\makeatletter
\providecommand \@ifxundefined [1]{%
 \@ifx{#1\undefined}
}%
\providecommand \@ifnum [1]{%
 \ifnum #1\expandafter \@firstoftwo
 \else \expandafter \@secondoftwo
 \fi
}%
\providecommand \@ifx [1]{%
 \ifx #1\expandafter \@firstoftwo
 \else \expandafter \@secondoftwo
 \fi
}%
\providecommand \natexlab [1]{#1}%
\providecommand \enquote  [1]{``#1''}%
\providecommand \bibnamefont  [1]{#1}%
\providecommand \bibfnamefont [1]{#1}%
\providecommand \citenamefont [1]{#1}%
\providecommand \href@noop [0]{\@secondoftwo}%
\providecommand \href [0]{\begingroup \@sanitize@url \@href}%
\providecommand \@href[1]{\@@startlink{#1}\@@href}%
\providecommand \@@href[1]{\endgroup#1\@@endlink}%
\providecommand \@sanitize@url [0]{\catcode `\\12\catcode `\$12\catcode
  `\&12\catcode `\#12\catcode `\^12\catcode `\_12\catcode `\%12\relax}%
\providecommand \@@startlink[1]{}%
\providecommand \@@endlink[0]{}%
\providecommand \url  [0]{\begingroup\@sanitize@url \@url }%
\providecommand \@url [1]{\endgroup\@href {#1}{\urlprefix }}%
\providecommand \urlprefix  [0]{URL }%
\providecommand \Eprint [0]{\href }%
\providecommand \doibase [0]{http://dx.doi.org/}%
\providecommand \selectlanguage [0]{\@gobble}%
\providecommand \bibinfo  [0]{\@secondoftwo}%
\providecommand \bibfield  [0]{\@secondoftwo}%
\providecommand \translation [1]{[#1]}%
\providecommand \BibitemOpen [0]{}%
\providecommand \bibitemStop [0]{}%
\providecommand \bibitemNoStop [0]{.\EOS\space}%
\providecommand \EOS [0]{\spacefactor3000\relax}%
\providecommand \BibitemShut  [1]{\csname bibitem#1\endcsname}%
\let\auto@bib@innerbib\@empty
\bibitem [{\citenamefont {Emma}\ \emph {et~al.}(2010)\citenamefont {Emma},
  \citenamefont {Akre}, \citenamefont {Arthur}, \citenamefont {Bionta},
  \citenamefont {Bostedt}, \citenamefont {Bozek}, \citenamefont {Brachmann},
  \citenamefont {Bucksbaum}, \citenamefont {Coffee}, \citenamefont {Decker},
  \citenamefont {Ding}, \citenamefont {Dowell}, \citenamefont {Edstrom},
  \citenamefont {Fisher}, \citenamefont {Frisch}, \citenamefont {Gilevich},
  \citenamefont {Hastings}, \citenamefont {Hays}, \citenamefont {Hering},
  \citenamefont {Huang}, \citenamefont {Iverson}, \citenamefont {Loos},
  \citenamefont {Messerschmidt}, \citenamefont {Miahnahri}, \citenamefont
  {Moeller}, \citenamefont {Nuhn}, \citenamefont {Pile}, \citenamefont
  {Ratner}, \citenamefont {Rzepiela}, \citenamefont {Schultz}, \citenamefont
  {Smith}, \citenamefont {Stefan}, \citenamefont {Tompkins}, \citenamefont
  {Turner}, \citenamefont {Welch}, \citenamefont {White}, \citenamefont {Wu},
  \citenamefont {Yocky},\ and\ \citenamefont {Galayda}}]{emma_first_2010}%
  \BibitemOpen
  \bibfield  {author} {\bibinfo {author} {\bibfnamefont {P.}~\bibnamefont
  {Emma}}, \bibinfo {author} {\bibfnamefont {R.}~\bibnamefont {Akre}}, \bibinfo
  {author} {\bibfnamefont {J.}~\bibnamefont {Arthur}}, \bibinfo {author}
  {\bibfnamefont {R.}~\bibnamefont {Bionta}}, \bibinfo {author} {\bibfnamefont
  {C.}~\bibnamefont {Bostedt}}, \bibinfo {author} {\bibfnamefont
  {J.}~\bibnamefont {Bozek}}, \bibinfo {author} {\bibfnamefont
  {A.}~\bibnamefont {Brachmann}}, \bibinfo {author} {\bibfnamefont
  {P.}~\bibnamefont {Bucksbaum}}, \bibinfo {author} {\bibfnamefont
  {R.}~\bibnamefont {Coffee}}, \bibinfo {author} {\bibfnamefont {F.-J.}\
  \bibnamefont {Decker}}, \bibinfo {author} {\bibfnamefont {Y.}~\bibnamefont
  {Ding}}, \bibinfo {author} {\bibfnamefont {D.}~\bibnamefont {Dowell}},
  \bibinfo {author} {\bibfnamefont {S.}~\bibnamefont {Edstrom}}, \bibinfo
  {author} {\bibfnamefont {A.}~\bibnamefont {Fisher}}, \bibinfo {author}
  {\bibfnamefont {J.}~\bibnamefont {Frisch}}, \bibinfo {author} {\bibfnamefont
  {S.}~\bibnamefont {Gilevich}}, \bibinfo {author} {\bibfnamefont
  {J.}~\bibnamefont {Hastings}}, \bibinfo {author} {\bibfnamefont
  {G.}~\bibnamefont {Hays}}, \bibinfo {author} {\bibfnamefont {P.}~\bibnamefont
  {Hering}}, \bibinfo {author} {\bibfnamefont {Z.}~\bibnamefont {Huang}},
  \bibinfo {author} {\bibfnamefont {R.}~\bibnamefont {Iverson}}, \bibinfo
  {author} {\bibfnamefont {H.}~\bibnamefont {Loos}}, \bibinfo {author}
  {\bibfnamefont {M.}~\bibnamefont {Messerschmidt}}, \bibinfo {author}
  {\bibfnamefont {A.}~\bibnamefont {Miahnahri}}, \bibinfo {author}
  {\bibfnamefont {S.}~\bibnamefont {Moeller}}, \bibinfo {author} {\bibfnamefont
  {H.-D.}\ \bibnamefont {Nuhn}}, \bibinfo {author} {\bibfnamefont
  {G.}~\bibnamefont {Pile}}, \bibinfo {author} {\bibfnamefont {D.}~\bibnamefont
  {Ratner}}, \bibinfo {author} {\bibfnamefont {J.}~\bibnamefont {Rzepiela}},
  \bibinfo {author} {\bibfnamefont {D.}~\bibnamefont {Schultz}}, \bibinfo
  {author} {\bibfnamefont {T.}~\bibnamefont {Smith}}, \bibinfo {author}
  {\bibfnamefont {P.}~\bibnamefont {Stefan}}, \bibinfo {author} {\bibfnamefont
  {H.}~\bibnamefont {Tompkins}}, \bibinfo {author} {\bibfnamefont
  {J.}~\bibnamefont {Turner}}, \bibinfo {author} {\bibfnamefont
  {J.}~\bibnamefont {Welch}}, \bibinfo {author} {\bibfnamefont
  {W.}~\bibnamefont {White}}, \bibinfo {author} {\bibfnamefont
  {J.}~\bibnamefont {Wu}}, \bibinfo {author} {\bibfnamefont {G.}~\bibnamefont
  {Yocky}}, \ and\ \bibinfo {author} {\bibfnamefont {J.}~\bibnamefont
  {Galayda}},\ }\bibfield  {title} {\enquote {\bibinfo {title} {First lasing
  and operation of an \aa ngstrom-wavelength free-electron laser},}\ }\href
  {\doibase 10.1038/nphoton.2010.176} {\bibfield  {journal} {\bibinfo
  {journal} {Nature Photonics}\ }\textbf {\bibinfo {volume} {4}},\ \bibinfo
  {pages} {641--647} (\bibinfo {year} {2010})}\BibitemShut {NoStop}%
\bibitem [{\citenamefont {Pellegrini}, \citenamefont {Marinelli},\ and\
  \citenamefont {Reiche}(2016)}]{pellegrini_physics_2016}%
  \BibitemOpen
  \bibfield  {author} {\bibinfo {author} {\bibfnamefont {C.}~\bibnamefont
  {Pellegrini}}, \bibinfo {author} {\bibfnamefont {A.}~\bibnamefont
  {Marinelli}}, \ and\ \bibinfo {author} {\bibfnamefont {S.}~\bibnamefont
  {Reiche}},\ }\bibfield  {title} {\enquote {\bibinfo {title} {The physics of
  x-ray free-electron lasers},}\ }\href {\doibase 10.1103/RevModPhys.88.015006}
  {\bibfield  {journal} {\bibinfo  {journal} {Reviews of Modern Physics}\
  }\textbf {\bibinfo {volume} {88}},\ \bibinfo {pages} {015006} (\bibinfo
  {year} {2016})}\BibitemShut {NoStop}%
\bibitem [{\citenamefont {Powers}\ \emph {et~al.}(2014)\citenamefont {Powers},
  \citenamefont {Ghebregziabher}, \citenamefont {Golovin}, \citenamefont {Liu},
  \citenamefont {Chen}, \citenamefont {Banerjee}, \citenamefont {Zhang},\ and\
  \citenamefont {Umstadter}}]{powers_quasi-monoenergetic_2014}%
  \BibitemOpen
  \bibfield  {author} {\bibinfo {author} {\bibfnamefont {N.~D.}\ \bibnamefont
  {Powers}}, \bibinfo {author} {\bibfnamefont {I.}~\bibnamefont
  {Ghebregziabher}}, \bibinfo {author} {\bibfnamefont {G.}~\bibnamefont
  {Golovin}}, \bibinfo {author} {\bibfnamefont {C.}~\bibnamefont {Liu}},
  \bibinfo {author} {\bibfnamefont {S.}~\bibnamefont {Chen}}, \bibinfo {author}
  {\bibfnamefont {S.}~\bibnamefont {Banerjee}}, \bibinfo {author}
  {\bibfnamefont {J.}~\bibnamefont {Zhang}}, \ and\ \bibinfo {author}
  {\bibfnamefont {D.~P.}\ \bibnamefont {Umstadter}},\ }\bibfield  {title}
  {\enquote {\bibinfo {title} {Quasi-monoenergetic and tunable {{X}}-rays from
  a laser-driven {{Compton}} light source},}\ }\href {\doibase
  10.1038/nphoton.2013.314} {\bibfield  {journal} {\bibinfo  {journal} {Nature
  Photonics}\ }\textbf {\bibinfo {volume} {8}},\ \bibinfo {pages} {28--31}
  (\bibinfo {year} {2014})}\BibitemShut {NoStop}%
\bibitem [{\citenamefont {Haynam}\ \emph {et~al.}(2007)\citenamefont {Haynam},
  \citenamefont {Wegner}, \citenamefont {Auerbach}, \citenamefont {Bowers},
  \citenamefont {Dixit}, \citenamefont {Erbert}, \citenamefont {Heestand},
  \citenamefont {Henesian}, \citenamefont {Hermann}, \citenamefont {Jancaitis},
  \citenamefont {Manes}, \citenamefont {Marshall}, \citenamefont {Mehta},
  \citenamefont {Menapace}, \citenamefont {Moses}, \citenamefont {Murray},
  \citenamefont {Nostrand}, \citenamefont {Orth}, \citenamefont {Patterson},
  \citenamefont {Sacks}, \citenamefont {Shaw}, \citenamefont {Spaeth},
  \citenamefont {Sutton}, \citenamefont {Williams}, \citenamefont {Widmayer},
  \citenamefont {White}, \citenamefont {Yang},\ and\ \citenamefont
  {Wonterghem}}]{haynam_national_2007}%
  \BibitemOpen
  \bibfield  {author} {\bibinfo {author} {\bibfnamefont {C.~A.}\ \bibnamefont
  {Haynam}}, \bibinfo {author} {\bibfnamefont {P.~J.}\ \bibnamefont {Wegner}},
  \bibinfo {author} {\bibfnamefont {J.~M.}\ \bibnamefont {Auerbach}}, \bibinfo
  {author} {\bibfnamefont {M.~W.}\ \bibnamefont {Bowers}}, \bibinfo {author}
  {\bibfnamefont {S.~N.}\ \bibnamefont {Dixit}}, \bibinfo {author}
  {\bibfnamefont {G.~V.}\ \bibnamefont {Erbert}}, \bibinfo {author}
  {\bibfnamefont {G.~M.}\ \bibnamefont {Heestand}}, \bibinfo {author}
  {\bibfnamefont {M.~A.}\ \bibnamefont {Henesian}}, \bibinfo {author}
  {\bibfnamefont {M.~R.}\ \bibnamefont {Hermann}}, \bibinfo {author}
  {\bibfnamefont {K.~S.}\ \bibnamefont {Jancaitis}}, \bibinfo {author}
  {\bibfnamefont {K.~R.}\ \bibnamefont {Manes}}, \bibinfo {author}
  {\bibfnamefont {C.~D.}\ \bibnamefont {Marshall}}, \bibinfo {author}
  {\bibfnamefont {N.~C.}\ \bibnamefont {Mehta}}, \bibinfo {author}
  {\bibfnamefont {J.}~\bibnamefont {Menapace}}, \bibinfo {author}
  {\bibfnamefont {E.}~\bibnamefont {Moses}}, \bibinfo {author} {\bibfnamefont
  {J.~R.}\ \bibnamefont {Murray}}, \bibinfo {author} {\bibfnamefont {M.~C.}\
  \bibnamefont {Nostrand}}, \bibinfo {author} {\bibfnamefont {C.~D.}\
  \bibnamefont {Orth}}, \bibinfo {author} {\bibfnamefont {R.}~\bibnamefont
  {Patterson}}, \bibinfo {author} {\bibfnamefont {R.~A.}\ \bibnamefont
  {Sacks}}, \bibinfo {author} {\bibfnamefont {M.~J.}\ \bibnamefont {Shaw}},
  \bibinfo {author} {\bibfnamefont {M.}~\bibnamefont {Spaeth}}, \bibinfo
  {author} {\bibfnamefont {S.~B.}\ \bibnamefont {Sutton}}, \bibinfo {author}
  {\bibfnamefont {W.~H.}\ \bibnamefont {Williams}}, \bibinfo {author}
  {\bibfnamefont {C.~C.}\ \bibnamefont {Widmayer}}, \bibinfo {author}
  {\bibfnamefont {R.~K.}\ \bibnamefont {White}}, \bibinfo {author}
  {\bibfnamefont {S.~T.}\ \bibnamefont {Yang}}, \ and\ \bibinfo {author}
  {\bibfnamefont {B.~M.~V.}\ \bibnamefont {Wonterghem}},\ }\bibfield  {title}
  {\enquote {\bibinfo {title} {National {{Ignition Facility}} laser performance
  status},}\ }\href {\doibase 10.1364/AO.46.003276} {\bibfield  {journal}
  {\bibinfo  {journal} {Applied Optics}\ }\textbf {\bibinfo {volume} {46}},\
  \bibinfo {pages} {3276--3303} (\bibinfo {year} {2007})}\BibitemShut {NoStop}%
\bibitem [{\citenamefont {Ferray}\ \emph {et~al.}(1988)\citenamefont {Ferray},
  \citenamefont {L'Huillier}, \citenamefont {Li}, \citenamefont {Lompre},
  \citenamefont {Mainfray},\ and\ \citenamefont
  {Manus}}]{ferray_multiple-harmonic_1988}%
  \BibitemOpen
  \bibfield  {author} {\bibinfo {author} {\bibfnamefont {M.}~\bibnamefont
  {Ferray}}, \bibinfo {author} {\bibfnamefont {A.}~\bibnamefont {L'Huillier}},
  \bibinfo {author} {\bibfnamefont {X.~F.}\ \bibnamefont {Li}}, \bibinfo
  {author} {\bibfnamefont {L.~A.}\ \bibnamefont {Lompre}}, \bibinfo {author}
  {\bibfnamefont {G.}~\bibnamefont {Mainfray}}, \ and\ \bibinfo {author}
  {\bibfnamefont {C.}~\bibnamefont {Manus}},\ }\bibfield  {title} {\enquote
  {\bibinfo {title} {Multiple-harmonic conversion of 1064 nm radiation in rare
  gases},}\ }\href {\doibase 10.1088/0953-4075/21/3/001} {\bibfield  {journal}
  {\bibinfo  {journal} {Journal of Physics B: Atomic, Molecular and Optical
  Physics}\ }\textbf {\bibinfo {volume} {21}},\ \bibinfo {pages} {L31--L35}
  (\bibinfo {year} {1988})}\BibitemShut {NoStop}%
\bibitem [{\citenamefont {Krausz}\ and\ \citenamefont
  {Ivanov}(2009)}]{krausz_attosecond_2009}%
  \BibitemOpen
  \bibfield  {author} {\bibinfo {author} {\bibfnamefont {F.}~\bibnamefont
  {Krausz}}\ and\ \bibinfo {author} {\bibfnamefont {M.}~\bibnamefont
  {Ivanov}},\ }\bibfield  {title} {\enquote {\bibinfo {title} {Attosecond
  physics},}\ }\href {\doibase 10.1103/RevModPhys.81.163} {\bibfield  {journal}
  {\bibinfo  {journal} {Reviews of Modern Physics}\ }\textbf {\bibinfo {volume}
  {81}},\ \bibinfo {pages} {163--234} (\bibinfo {year} {2009})}\BibitemShut
  {NoStop}%
\bibitem [{\citenamefont {Popmintchev}\ \emph {et~al.}(2012)\citenamefont
  {Popmintchev}, \citenamefont {Chen}, \citenamefont {Popmintchev},
  \citenamefont {Arpin}, \citenamefont {Brown}, \citenamefont {Ali{\v
  s}auskas}, \citenamefont {Andriukaitis}, \citenamefont {Bal{\v c}iunas},
  \citenamefont {M{\"u}cke}, \citenamefont {Pugzlys}, \citenamefont {Baltu{\v
  s}ka}, \citenamefont {Shim}, \citenamefont {Schrauth}, \citenamefont {Gaeta},
  \citenamefont {{Hern{\'a}ndez-Garc{\'i}a}}, \citenamefont {Plaja},
  \citenamefont {Becker}, \citenamefont {{Jaron-Becker}}, \citenamefont
  {Murnane},\ and\ \citenamefont {Kapteyn}}]{popmintchev_bright_2012}%
  \BibitemOpen
  \bibfield  {author} {\bibinfo {author} {\bibfnamefont {T.}~\bibnamefont
  {Popmintchev}}, \bibinfo {author} {\bibfnamefont {M.-C.}\ \bibnamefont
  {Chen}}, \bibinfo {author} {\bibfnamefont {D.}~\bibnamefont {Popmintchev}},
  \bibinfo {author} {\bibfnamefont {P.}~\bibnamefont {Arpin}}, \bibinfo
  {author} {\bibfnamefont {S.}~\bibnamefont {Brown}}, \bibinfo {author}
  {\bibfnamefont {S.}~\bibnamefont {Ali{\v s}auskas}}, \bibinfo {author}
  {\bibfnamefont {G.}~\bibnamefont {Andriukaitis}}, \bibinfo {author}
  {\bibfnamefont {T.}~\bibnamefont {Bal{\v c}iunas}}, \bibinfo {author}
  {\bibfnamefont {O.~D.}\ \bibnamefont {M{\"u}cke}}, \bibinfo {author}
  {\bibfnamefont {A.}~\bibnamefont {Pugzlys}}, \bibinfo {author} {\bibfnamefont
  {A.}~\bibnamefont {Baltu{\v s}ka}}, \bibinfo {author} {\bibfnamefont
  {B.}~\bibnamefont {Shim}}, \bibinfo {author} {\bibfnamefont {S.~E.}\
  \bibnamefont {Schrauth}}, \bibinfo {author} {\bibfnamefont {A.}~\bibnamefont
  {Gaeta}}, \bibinfo {author} {\bibfnamefont {C.}~\bibnamefont
  {{Hern{\'a}ndez-Garc{\'i}a}}}, \bibinfo {author} {\bibfnamefont
  {L.}~\bibnamefont {Plaja}}, \bibinfo {author} {\bibfnamefont
  {A.}~\bibnamefont {Becker}}, \bibinfo {author} {\bibfnamefont
  {A.}~\bibnamefont {{Jaron-Becker}}}, \bibinfo {author} {\bibfnamefont
  {M.~M.}\ \bibnamefont {Murnane}}, \ and\ \bibinfo {author} {\bibfnamefont
  {H.~C.}\ \bibnamefont {Kapteyn}},\ }\bibfield  {title} {\enquote {\bibinfo
  {title} {Bright {{Coherent Ultrahigh Harmonics}} in the {{keV X}}-ray
  {{Regime}} from {{Mid}}-{{Infrared Femtosecond Lasers}}},}\ }\href {\doibase
  10.1126/science.1218497} {\bibfield  {journal} {\bibinfo  {journal}
  {Science}\ }\textbf {\bibinfo {volume} {336}},\ \bibinfo {pages} {1287--1291}
  (\bibinfo {year} {2012})}\BibitemShut {NoStop}%
\bibitem [{\citenamefont {Ghimire}\ \emph {et~al.}(2011)\citenamefont
  {Ghimire}, \citenamefont {DiChiara}, \citenamefont {Sistrunk}, \citenamefont
  {Agostini}, \citenamefont {DiMauro},\ and\ \citenamefont
  {Reis}}]{ghimire_observation_2011}%
  \BibitemOpen
  \bibfield  {author} {\bibinfo {author} {\bibfnamefont {S.}~\bibnamefont
  {Ghimire}}, \bibinfo {author} {\bibfnamefont {A.~D.}\ \bibnamefont
  {DiChiara}}, \bibinfo {author} {\bibfnamefont {E.}~\bibnamefont {Sistrunk}},
  \bibinfo {author} {\bibfnamefont {P.}~\bibnamefont {Agostini}}, \bibinfo
  {author} {\bibfnamefont {L.~F.}\ \bibnamefont {DiMauro}}, \ and\ \bibinfo
  {author} {\bibfnamefont {D.~A.}\ \bibnamefont {Reis}},\ }\bibfield  {title}
  {\enquote {\bibinfo {title} {Observation of high-order harmonic generation in
  a bulk crystal},}\ }\href {\doibase 10.1038/nphys1847} {\bibfield  {journal}
  {\bibinfo  {journal} {Nature Physics}\ }\textbf {\bibinfo {volume} {7}},\
  \bibinfo {pages} {138--141} (\bibinfo {year} {2011})}\BibitemShut {NoStop}%
\bibitem [{\citenamefont {Ghimire}\ and\ \citenamefont
  {Reis}(2019)}]{ghimire_high-harmonic_2019}%
  \BibitemOpen
  \bibfield  {author} {\bibinfo {author} {\bibfnamefont {S.}~\bibnamefont
  {Ghimire}}\ and\ \bibinfo {author} {\bibfnamefont {D.~A.}\ \bibnamefont
  {Reis}},\ }\bibfield  {title} {\enquote {\bibinfo {title} {High-harmonic
  generation from solids},}\ }\href {\doibase 10.1038/s41567-018-0315-5}
  {\bibfield  {journal} {\bibinfo  {journal} {Nature Physics}\ }\textbf
  {\bibinfo {volume} {15}},\ \bibinfo {pages} {10--16} (\bibinfo {year}
  {2019})}\BibitemShut {NoStop}%
\bibitem [{\citenamefont {Teubner}\ and\ \citenamefont
  {Gibbon}(2009)}]{RevModPhys.81.445}%
  \BibitemOpen
  \bibfield  {author} {\bibinfo {author} {\bibfnamefont {U.}~\bibnamefont
  {Teubner}}\ and\ \bibinfo {author} {\bibfnamefont {P.}~\bibnamefont
  {Gibbon}},\ }\bibfield  {title} {\enquote {\bibinfo {title} {High-order
  harmonics from laser-irradiated plasma surfaces},}\ }\href {\doibase
  10.1103/RevModPhys.81.445} {\bibfield  {journal} {\bibinfo  {journal} {Rev.
  Mod. Phys.}\ }\textbf {\bibinfo {volume} {81}},\ \bibinfo {pages} {445--479}
  (\bibinfo {year} {2009})}\BibitemShut {NoStop}%
\bibitem [{\citenamefont {Malkin}, \citenamefont {Shvets},\ and\ \citenamefont
  {Fisch}(1999)}]{malkin_fast_1999-1}%
  \BibitemOpen
  \bibfield  {author} {\bibinfo {author} {\bibfnamefont {V.~M.}\ \bibnamefont
  {Malkin}}, \bibinfo {author} {\bibfnamefont {G.}~\bibnamefont {Shvets}}, \
  and\ \bibinfo {author} {\bibfnamefont {N.~J.}\ \bibnamefont {Fisch}},\
  }\bibfield  {title} {\enquote {\bibinfo {title} {Fast {{Compression}} of
  {{Laser Beams}} to {{Highly Overcritical Powers}}},}\ }\href {\doibase
  10.1103/PhysRevLett.82.4448} {\bibfield  {journal} {\bibinfo  {journal}
  {Physical Review Letters}\ }\textbf {\bibinfo {volume} {82}},\ \bibinfo
  {pages} {4448--4451} (\bibinfo {year} {1999})}\BibitemShut {NoStop}%
\bibitem [{\citenamefont {Ping}\ \emph {et~al.}(2004)\citenamefont {Ping},
  \citenamefont {Cheng}, \citenamefont {Suckewer}, \citenamefont {Clark},\ and\
  \citenamefont {Fisch}}]{Ping2004}%
  \BibitemOpen
  \bibfield  {author} {\bibinfo {author} {\bibfnamefont {Y.}~\bibnamefont
  {Ping}}, \bibinfo {author} {\bibfnamefont {W.}~\bibnamefont {Cheng}},
  \bibinfo {author} {\bibfnamefont {S.}~\bibnamefont {Suckewer}}, \bibinfo
  {author} {\bibfnamefont {D.~S.}\ \bibnamefont {Clark}}, \ and\ \bibinfo
  {author} {\bibfnamefont {N.~J.}\ \bibnamefont {Fisch}},\ }\bibfield  {title}
  {\enquote {\bibinfo {title} {{Amplification of ultrashort laser pulses by a
  resonant Raman scheme in a gas-jet plasma}},}\ }\href@noop {} {\bibfield
  {journal} {\bibinfo  {journal} {Phys. Rev. Lett.}\ }\textbf {\bibinfo
  {volume} {92}},\ \bibinfo {pages} {175007} (\bibinfo {year}
  {2004})}\BibitemShut {NoStop}%
\bibitem [{\citenamefont {Ren}\ \emph {et~al.}(2008)\citenamefont {Ren},
  \citenamefont {Li}, \citenamefont {Morozov}, \citenamefont {Suckewer},
  \citenamefont {Yampolsky}, \citenamefont {Malkin},\ and\ \citenamefont
  {Fisch}}]{ren_compact_2008}%
  \BibitemOpen
  \bibfield  {author} {\bibinfo {author} {\bibfnamefont {J.}~\bibnamefont
  {Ren}}, \bibinfo {author} {\bibfnamefont {S.}~\bibnamefont {Li}}, \bibinfo
  {author} {\bibfnamefont {A.}~\bibnamefont {Morozov}}, \bibinfo {author}
  {\bibfnamefont {S.}~\bibnamefont {Suckewer}}, \bibinfo {author}
  {\bibfnamefont {N.~A.}\ \bibnamefont {Yampolsky}}, \bibinfo {author}
  {\bibfnamefont {V.~M.}\ \bibnamefont {Malkin}}, \ and\ \bibinfo {author}
  {\bibfnamefont {N.~J.}\ \bibnamefont {Fisch}},\ }\bibfield  {title} {\enquote
  {\bibinfo {title} {A compact double-pass {{Raman}} backscattering
  amplifier/compressor},}\ }\href {\doibase 10.1063/1.2844352} {\bibfield
  {journal} {\bibinfo  {journal} {Physics of Plasmas}\ }\textbf {\bibinfo
  {volume} {15}},\ \bibinfo {pages} {056702} (\bibinfo {year}
  {2008})}\BibitemShut {NoStop}%
\bibitem [{\citenamefont {Pai}\ \emph {et~al.}(2008)\citenamefont {Pai},
  \citenamefont {Lin}, \citenamefont {Ha}, \citenamefont {Huang}, \citenamefont
  {Tsou}, \citenamefont {Chu}, \citenamefont {Lin}, \citenamefont {Wang},\ and\
  \citenamefont {Chen}}]{pai_backward_2008}%
  \BibitemOpen
  \bibfield  {author} {\bibinfo {author} {\bibfnamefont {C.-H.}\ \bibnamefont
  {Pai}}, \bibinfo {author} {\bibfnamefont {M.-W.}\ \bibnamefont {Lin}},
  \bibinfo {author} {\bibfnamefont {L.-C.}\ \bibnamefont {Ha}}, \bibinfo
  {author} {\bibfnamefont {S.-T.}\ \bibnamefont {Huang}}, \bibinfo {author}
  {\bibfnamefont {Y.-C.}\ \bibnamefont {Tsou}}, \bibinfo {author}
  {\bibfnamefont {H.-H.}\ \bibnamefont {Chu}}, \bibinfo {author} {\bibfnamefont
  {J.-Y.}\ \bibnamefont {Lin}}, \bibinfo {author} {\bibfnamefont
  {J.}~\bibnamefont {Wang}}, \ and\ \bibinfo {author} {\bibfnamefont {S.-Y.}\
  \bibnamefont {Chen}},\ }\bibfield  {title} {\enquote {\bibinfo {title}
  {Backward {{Raman Amplification}} in a {{Plasma Waveguide}}},}\ }\href
  {\doibase 10.1103/PhysRevLett.101.065005} {\bibfield  {journal} {\bibinfo
  {journal} {Physical Review Letters}\ }\textbf {\bibinfo {volume} {101}},\
  \bibinfo {pages} {065005} (\bibinfo {year} {2008})}\BibitemShut {NoStop}%
\bibitem [{\citenamefont {Ping}\ \emph {et~al.}(2009)\citenamefont {Ping},
  \citenamefont {Kirkwood}, \citenamefont {Wang}, \citenamefont {Clark},
  \citenamefont {Wilks}, \citenamefont {Meezan}, \citenamefont {Berger},
  \citenamefont {Wurtele}, \citenamefont {Fisch}, \citenamefont {Malkin},
  \citenamefont {Valeo}, \citenamefont {Martins},\ and\ \citenamefont
  {Joshi}}]{ping_development_2009}%
  \BibitemOpen
  \bibfield  {author} {\bibinfo {author} {\bibfnamefont {Y.}~\bibnamefont
  {Ping}}, \bibinfo {author} {\bibfnamefont {R.~K.}\ \bibnamefont {Kirkwood}},
  \bibinfo {author} {\bibfnamefont {T.-L.}\ \bibnamefont {Wang}}, \bibinfo
  {author} {\bibfnamefont {D.~S.}\ \bibnamefont {Clark}}, \bibinfo {author}
  {\bibfnamefont {S.~C.}\ \bibnamefont {Wilks}}, \bibinfo {author}
  {\bibfnamefont {N.}~\bibnamefont {Meezan}}, \bibinfo {author} {\bibfnamefont
  {R.~L.}\ \bibnamefont {Berger}}, \bibinfo {author} {\bibfnamefont
  {J.}~\bibnamefont {Wurtele}}, \bibinfo {author} {\bibfnamefont {N.~J.}\
  \bibnamefont {Fisch}}, \bibinfo {author} {\bibfnamefont {V.~M.}\ \bibnamefont
  {Malkin}}, \bibinfo {author} {\bibfnamefont {E.~J.}\ \bibnamefont {Valeo}},
  \bibinfo {author} {\bibfnamefont {S.~F.}\ \bibnamefont {Martins}}, \ and\
  \bibinfo {author} {\bibfnamefont {C.}~\bibnamefont {Joshi}},\ }\bibfield
  {title} {\enquote {\bibinfo {title} {Development of a nanosecond-laser-pumped
  {{Raman}} amplifier for short laser pulses in plasma},}\ }\href {\doibase
  10.1063/1.3276739} {\bibfield  {journal} {\bibinfo  {journal} {Physics of
  Plasmas}\ }\textbf {\bibinfo {volume} {16}},\ \bibinfo {pages} {123113}
  (\bibinfo {year} {2009})}\BibitemShut {NoStop}%
\bibitem [{\citenamefont {Trines}\ \emph {et~al.}(2011)\citenamefont {Trines},
  \citenamefont {Fi{\'u}za}, \citenamefont {Bingham}, \citenamefont {Fonseca},
  \citenamefont {Silva}, \citenamefont {Cairns},\ and\ \citenamefont
  {Norreys}}]{trines_simulations_2011}%
  \BibitemOpen
  \bibfield  {author} {\bibinfo {author} {\bibfnamefont {R.~M. G.~M.}\
  \bibnamefont {Trines}}, \bibinfo {author} {\bibfnamefont {F.}~\bibnamefont
  {Fi{\'u}za}}, \bibinfo {author} {\bibfnamefont {R.}~\bibnamefont {Bingham}},
  \bibinfo {author} {\bibfnamefont {R.~A.}\ \bibnamefont {Fonseca}}, \bibinfo
  {author} {\bibfnamefont {L.~O.}\ \bibnamefont {Silva}}, \bibinfo {author}
  {\bibfnamefont {R.~A.}\ \bibnamefont {Cairns}}, \ and\ \bibinfo {author}
  {\bibfnamefont {P.~A.}\ \bibnamefont {Norreys}},\ }\bibfield  {title}
  {\enquote {\bibinfo {title} {Simulations of efficient {{Raman}} amplification
  into the multipetawatt regime},}\ }\href {\doibase 10.1038/nphys1793}
  {\bibfield  {journal} {\bibinfo  {journal} {Nature Physics}\ }\textbf
  {\bibinfo {volume} {7}},\ \bibinfo {pages} {87--92} (\bibinfo {year}
  {2011})}\BibitemShut {NoStop}%
\bibitem [{\citenamefont {Vieux}\ \emph {et~al.}(2011)\citenamefont {Vieux},
  \citenamefont {Lyachev}, \citenamefont {Yang}, \citenamefont {Ersfeld},
  \citenamefont {Farmer}, \citenamefont {Brunetti}, \citenamefont {Issac},
  \citenamefont {Raj}, \citenamefont {Welsh}, \citenamefont {Wiggins} \emph
  {et~al.}}]{Vieux2011}%
  \BibitemOpen
  \bibfield  {author} {\bibinfo {author} {\bibfnamefont {G.}~\bibnamefont
  {Vieux}}, \bibinfo {author} {\bibfnamefont {A.}~\bibnamefont {Lyachev}},
  \bibinfo {author} {\bibfnamefont {X.}~\bibnamefont {Yang}}, \bibinfo {author}
  {\bibfnamefont {B.}~\bibnamefont {Ersfeld}}, \bibinfo {author} {\bibfnamefont
  {J.}~\bibnamefont {Farmer}}, \bibinfo {author} {\bibfnamefont
  {E.}~\bibnamefont {Brunetti}}, \bibinfo {author} {\bibfnamefont
  {R.}~\bibnamefont {Issac}}, \bibinfo {author} {\bibfnamefont
  {G.}~\bibnamefont {Raj}}, \bibinfo {author} {\bibfnamefont {G.}~\bibnamefont
  {Welsh}}, \bibinfo {author} {\bibfnamefont {S.}~\bibnamefont {Wiggins}},
  \emph {et~al.},\ }\bibfield  {title} {\enquote {\bibinfo {title} {{Chirped
  pulse Raman amplification in plasma}},}\ }\href@noop {} {\bibfield  {journal}
  {\bibinfo  {journal} {New J. Phys.}\ }\textbf {\bibinfo {volume} {13}},\
  \bibinfo {pages} {063042} (\bibinfo {year} {2011})}\BibitemShut {NoStop}%
\bibitem [{\citenamefont {Yampolsky}\ and\ \citenamefont
  {Fisch}(2011)}]{Yampolsky2011}%
  \BibitemOpen
  \bibfield  {author} {\bibinfo {author} {\bibfnamefont {N.~A.}\ \bibnamefont
  {Yampolsky}}\ and\ \bibinfo {author} {\bibfnamefont {N.~J.}\ \bibnamefont
  {Fisch}},\ }\bibfield  {title} {\enquote {\bibinfo {title} {{Limiting effects
  on laser compression by resonant backward Raman scattering in modern
  experiments}},}\ }\href@noop {} {\bibfield  {journal} {\bibinfo  {journal}
  {Phys. Plasmas}\ }\textbf {\bibinfo {volume} {18}},\ \bibinfo {pages}
  {056711} (\bibinfo {year} {2011})}\BibitemShut {NoStop}%
\bibitem [{\citenamefont {Vieux}\ \emph {et~al.}(2017)\citenamefont {Vieux},
  \citenamefont {Cipiccia}, \citenamefont {Grant}, \citenamefont {Lemos},
  \citenamefont {Grant}, \citenamefont {Ciocarlan}, \citenamefont {Ersfeld},
  \citenamefont {Hur}, \citenamefont {Lepipas}, \citenamefont {Manahan},
  \citenamefont {Raj}, \citenamefont {Reboredo~Gil}, \citenamefont {Subiel},
  \citenamefont {Welsh}, \citenamefont {Wiggins}, \citenamefont {Yoffe},
  \citenamefont {Farmer}, \citenamefont {Aniculaesei}, \citenamefont
  {Brunetti}, \citenamefont {Yang}, \citenamefont {Heathcote}, \citenamefont
  {Nersisyan}, \citenamefont {Lewis}, \citenamefont {Pukhov}, \citenamefont
  {Dias},\ and\ \citenamefont {Jaroszynski}}]{vieux_ultra-high_2017}%
  \BibitemOpen
  \bibfield  {author} {\bibinfo {author} {\bibfnamefont {G.}~\bibnamefont
  {Vieux}}, \bibinfo {author} {\bibfnamefont {S.}~\bibnamefont {Cipiccia}},
  \bibinfo {author} {\bibfnamefont {D.~W.}\ \bibnamefont {Grant}}, \bibinfo
  {author} {\bibfnamefont {N.}~\bibnamefont {Lemos}}, \bibinfo {author}
  {\bibfnamefont {P.}~\bibnamefont {Grant}}, \bibinfo {author} {\bibfnamefont
  {C.}~\bibnamefont {Ciocarlan}}, \bibinfo {author} {\bibfnamefont
  {B.}~\bibnamefont {Ersfeld}}, \bibinfo {author} {\bibfnamefont {M.~S.}\
  \bibnamefont {Hur}}, \bibinfo {author} {\bibfnamefont {P.}~\bibnamefont
  {Lepipas}}, \bibinfo {author} {\bibfnamefont {G.~G.}\ \bibnamefont
  {Manahan}}, \bibinfo {author} {\bibfnamefont {G.}~\bibnamefont {Raj}},
  \bibinfo {author} {\bibfnamefont {D.}~\bibnamefont {Reboredo~Gil}}, \bibinfo
  {author} {\bibfnamefont {A.}~\bibnamefont {Subiel}}, \bibinfo {author}
  {\bibfnamefont {G.~H.}\ \bibnamefont {Welsh}}, \bibinfo {author}
  {\bibfnamefont {S.~M.}\ \bibnamefont {Wiggins}}, \bibinfo {author}
  {\bibfnamefont {S.~R.}\ \bibnamefont {Yoffe}}, \bibinfo {author}
  {\bibfnamefont {J.~P.}\ \bibnamefont {Farmer}}, \bibinfo {author}
  {\bibfnamefont {C.}~\bibnamefont {Aniculaesei}}, \bibinfo {author}
  {\bibfnamefont {E.}~\bibnamefont {Brunetti}}, \bibinfo {author}
  {\bibfnamefont {X.}~\bibnamefont {Yang}}, \bibinfo {author} {\bibfnamefont
  {R.}~\bibnamefont {Heathcote}}, \bibinfo {author} {\bibfnamefont
  {G.}~\bibnamefont {Nersisyan}}, \bibinfo {author} {\bibfnamefont {C.~L.~S.}\
  \bibnamefont {Lewis}}, \bibinfo {author} {\bibfnamefont {A.}~\bibnamefont
  {Pukhov}}, \bibinfo {author} {\bibfnamefont {J.~M.}\ \bibnamefont {Dias}}, \
  and\ \bibinfo {author} {\bibfnamefont {D.~A.}\ \bibnamefont {Jaroszynski}},\
  }\bibfield  {title} {\enquote {\bibinfo {title} {An ultra-high gain and
  efficient amplifier based on {{Raman}} amplification in plasma},}\ }\href
  {\doibase 10.1038/s41598-017-01783-4} {\bibfield  {journal} {\bibinfo
  {journal} {Scientific Reports}\ }\textbf {\bibinfo {volume} {7}},\ \bibinfo
  {pages} {2399} (\bibinfo {year} {2017})}\BibitemShut {NoStop}%
\bibitem [{\citenamefont {Turnbull}\ \emph {et~al.}(2018)\citenamefont
  {Turnbull}, \citenamefont {Bucht}, \citenamefont {Davies}, \citenamefont
  {Haberberger}, \citenamefont {Kessler}, \citenamefont {Shaw},\ and\
  \citenamefont {Froula}}]{PhysRevLett.120.024801}%
  \BibitemOpen
  \bibfield  {author} {\bibinfo {author} {\bibfnamefont {D.}~\bibnamefont
  {Turnbull}}, \bibinfo {author} {\bibfnamefont {S.}~\bibnamefont {Bucht}},
  \bibinfo {author} {\bibfnamefont {A.}~\bibnamefont {Davies}}, \bibinfo
  {author} {\bibfnamefont {D.}~\bibnamefont {Haberberger}}, \bibinfo {author}
  {\bibfnamefont {T.}~\bibnamefont {Kessler}}, \bibinfo {author} {\bibfnamefont
  {J.~L.}\ \bibnamefont {Shaw}}, \ and\ \bibinfo {author} {\bibfnamefont
  {D.~H.}\ \bibnamefont {Froula}},\ }\bibfield  {title} {\enquote {\bibinfo
  {title} {{Raman Amplification with a Flying Focus}},}\ }\href {\doibase
  10.1103/PhysRevLett.120.024801} {\bibfield  {journal} {\bibinfo  {journal}
  {Phys. Rev. Lett.}\ }\textbf {\bibinfo {volume} {120}},\ \bibinfo {pages}
  {024801} (\bibinfo {year} {2018})}\BibitemShut {NoStop}%
\bibitem [{\citenamefont {Balakin}, \citenamefont {Levin},\ and\ \citenamefont
  {Skobelev}(2020)}]{Balakin2020}%
  \BibitemOpen
  \bibfield  {author} {\bibinfo {author} {\bibfnamefont {A.~A.}\ \bibnamefont
  {Balakin}}, \bibinfo {author} {\bibfnamefont {D.~S.}\ \bibnamefont {Levin}},
  \ and\ \bibinfo {author} {\bibfnamefont {S.~A.}\ \bibnamefont {Skobelev}},\
  }\bibfield  {title} {\enquote {\bibinfo {title} {{Compression of laser pulses
  due to Raman amplification of plasma noises}},}\ }\href {\doibase
  {10.1103/PhysRevA.102.013516}} {\bibfield  {journal} {\bibinfo  {journal}
  {{Physical Review A}}\ }\textbf {\bibinfo {volume} {{102}}},\ \bibinfo
  {pages} {{013516}} (\bibinfo {year} {{2020}})}\BibitemShut {NoStop}%
\bibitem [{\citenamefont {Lancia}\ \emph {et~al.}(2010)\citenamefont {Lancia},
  \citenamefont {Marqu\`es}, \citenamefont {Nakatsutsumi}, \citenamefont
  {Riconda}, \citenamefont {Weber}, \citenamefont {H\"uller}, \citenamefont
  {Man\ifmmode \check{c}\else \v{c}\fi{}i\ifmmode~\acute{c}\else \'{c}\fi{}},
  \citenamefont {Antici}, \citenamefont {Tikhonchuk}, \citenamefont {H\'eron},
  \citenamefont {Audebert},\ and\ \citenamefont {Fuchs}}]{Lancia2010}%
  \BibitemOpen
  \bibfield  {author} {\bibinfo {author} {\bibfnamefont {L.}~\bibnamefont
  {Lancia}}, \bibinfo {author} {\bibfnamefont {J.-R.}\ \bibnamefont
  {Marqu\`es}}, \bibinfo {author} {\bibfnamefont {M.}~\bibnamefont
  {Nakatsutsumi}}, \bibinfo {author} {\bibfnamefont {C.}~\bibnamefont
  {Riconda}}, \bibinfo {author} {\bibfnamefont {S.}~\bibnamefont {Weber}},
  \bibinfo {author} {\bibfnamefont {S.}~\bibnamefont {H\"uller}}, \bibinfo
  {author} {\bibfnamefont {A.}~\bibnamefont {Man\ifmmode \check{c}\else
  \v{c}\fi{}i\ifmmode~\acute{c}\else \'{c}\fi{}}}, \bibinfo {author}
  {\bibfnamefont {P.}~\bibnamefont {Antici}}, \bibinfo {author} {\bibfnamefont
  {V.~T.}\ \bibnamefont {Tikhonchuk}}, \bibinfo {author} {\bibfnamefont
  {A.}~\bibnamefont {H\'eron}}, \bibinfo {author} {\bibfnamefont
  {P.}~\bibnamefont {Audebert}}, \ and\ \bibinfo {author} {\bibfnamefont
  {J.}~\bibnamefont {Fuchs}},\ }\bibfield  {title} {\enquote {\bibinfo {title}
  {Experimental evidence of short light pulse amplification using
  strong-coupling stimulated brillouin scattering in the pump depletion
  regime},}\ }\href {\doibase 10.1103/PhysRevLett.104.025001} {\bibfield
  {journal} {\bibinfo  {journal} {Phys. Rev. Lett.}\ }\textbf {\bibinfo
  {volume} {104}},\ \bibinfo {pages} {025001} (\bibinfo {year}
  {2010})}\BibitemShut {NoStop}%
\bibitem [{\citenamefont {Lehmann}\ and\ \citenamefont
  {Spatschek}(2013)}]{lehmann_nonlinear_2013}%
  \BibitemOpen
  \bibfield  {author} {\bibinfo {author} {\bibfnamefont {G.}~\bibnamefont
  {Lehmann}}\ and\ \bibinfo {author} {\bibfnamefont {K.~H.}\ \bibnamefont
  {Spatschek}},\ }\bibfield  {title} {\enquote {\bibinfo {title} {Nonlinear
  {{Brillouin}} amplification of finite-duration seeds in the strong coupling
  regime},}\ }\href {\doibase 10.1063/1.4816030} {\bibfield  {journal}
  {\bibinfo  {journal} {Physics of Plasmas}\ }\textbf {\bibinfo {volume}
  {20}},\ \bibinfo {pages} {073112} (\bibinfo {year} {2013})}\BibitemShut
  {NoStop}%
\bibitem [{\citenamefont {Riconda}\ \emph {et~al.}(2013)\citenamefont
  {Riconda}, \citenamefont {Weber}, \citenamefont {Lancia}, \citenamefont
  {Marqu{\`e}s}, \citenamefont {Mourou},\ and\ \citenamefont
  {Fuchs}}]{riconda_spectral_2013-1}%
  \BibitemOpen
  \bibfield  {author} {\bibinfo {author} {\bibfnamefont {C.}~\bibnamefont
  {Riconda}}, \bibinfo {author} {\bibfnamefont {S.}~\bibnamefont {Weber}},
  \bibinfo {author} {\bibfnamefont {L.}~\bibnamefont {Lancia}}, \bibinfo
  {author} {\bibfnamefont {J.-R.}\ \bibnamefont {Marqu{\`e}s}}, \bibinfo
  {author} {\bibfnamefont {G.~A.}\ \bibnamefont {Mourou}}, \ and\ \bibinfo
  {author} {\bibfnamefont {J.}~\bibnamefont {Fuchs}},\ }\bibfield  {title}
  {\enquote {\bibinfo {title} {Spectral characteristics of ultra-short laser
  pulses in plasma amplifiers},}\ }\href {\doibase 10.1063/1.4818893}
  {\bibfield  {journal} {\bibinfo  {journal} {Physics of Plasmas}\ }\textbf
  {\bibinfo {volume} {20}},\ \bibinfo {pages} {083115} (\bibinfo {year}
  {2013})}\BibitemShut {NoStop}%
\bibitem [{\citenamefont {Edwards}, \citenamefont {Mikhailova},\ and\
  \citenamefont {Fisch}(2017)}]{edwards_x-ray_2017}%
  \BibitemOpen
  \bibfield  {author} {\bibinfo {author} {\bibfnamefont {M.~R.}\ \bibnamefont
  {Edwards}}, \bibinfo {author} {\bibfnamefont {J.~M.}\ \bibnamefont
  {Mikhailova}}, \ and\ \bibinfo {author} {\bibfnamefont {N.~J.}\ \bibnamefont
  {Fisch}},\ }\bibfield  {title} {\enquote {\bibinfo {title} {X-ray
  amplification by stimulated {{Brillouin}} scattering},}\ }\href {\doibase
  10.1103/PhysRevE.96.023209} {\bibfield  {journal} {\bibinfo  {journal}
  {Physical Review E}\ }\textbf {\bibinfo {volume} {96}},\ \bibinfo {pages}
  {023209} (\bibinfo {year} {2017})}\BibitemShut {NoStop}%
\bibitem [{\citenamefont {Kirkwood}\ \emph {et~al.}(2018)\citenamefont
  {Kirkwood}, \citenamefont {Turnbull}, \citenamefont {Chapman}, \citenamefont
  {Wilks}, \citenamefont {Rosen}, \citenamefont {London}, \citenamefont
  {Pickworth}, \citenamefont {Dunlop}, \citenamefont {Moody}, \citenamefont
  {Strozzi}, \citenamefont {Michel}, \citenamefont {Divol}, \citenamefont
  {Landen}, \citenamefont {MacGowan}, \citenamefont {Van~Wonterghem},
  \citenamefont {Fournier},\ and\ \citenamefont
  {Blue}}]{kirkwood_plasma-based_2018}%
  \BibitemOpen
  \bibfield  {author} {\bibinfo {author} {\bibfnamefont {R.~K.}\ \bibnamefont
  {Kirkwood}}, \bibinfo {author} {\bibfnamefont {D.~P.}\ \bibnamefont
  {Turnbull}}, \bibinfo {author} {\bibfnamefont {T.}~\bibnamefont {Chapman}},
  \bibinfo {author} {\bibfnamefont {S.~C.}\ \bibnamefont {Wilks}}, \bibinfo
  {author} {\bibfnamefont {M.~D.}\ \bibnamefont {Rosen}}, \bibinfo {author}
  {\bibfnamefont {R.~A.}\ \bibnamefont {London}}, \bibinfo {author}
  {\bibfnamefont {L.~A.}\ \bibnamefont {Pickworth}}, \bibinfo {author}
  {\bibfnamefont {W.~H.}\ \bibnamefont {Dunlop}}, \bibinfo {author}
  {\bibfnamefont {J.~D.}\ \bibnamefont {Moody}}, \bibinfo {author}
  {\bibfnamefont {D.~J.}\ \bibnamefont {Strozzi}}, \bibinfo {author}
  {\bibfnamefont {P.~A.}\ \bibnamefont {Michel}}, \bibinfo {author}
  {\bibfnamefont {L.}~\bibnamefont {Divol}}, \bibinfo {author} {\bibfnamefont
  {O.~L.}\ \bibnamefont {Landen}}, \bibinfo {author} {\bibfnamefont {B.~J.}\
  \bibnamefont {MacGowan}}, \bibinfo {author} {\bibfnamefont {B.~M.}\
  \bibnamefont {Van~Wonterghem}}, \bibinfo {author} {\bibfnamefont {K.~B.}\
  \bibnamefont {Fournier}}, \ and\ \bibinfo {author} {\bibfnamefont {B.~E.}\
  \bibnamefont {Blue}},\ }\bibfield  {title} {\enquote {\bibinfo {title}
  {Plasma-based beam combiner for very high fluence and energy},}\ }\href
  {\doibase 10.1038/nphys4271} {\bibfield  {journal} {\bibinfo  {journal}
  {Nature Physics}\ }\textbf {\bibinfo {volume} {14}},\ \bibinfo {pages}
  {80--84} (\bibinfo {year} {2018})}\BibitemShut {NoStop}%
\bibitem [{\citenamefont {Shi}, \citenamefont {Qin},\ and\ \citenamefont
  {Fisch}(2018)}]{Shi2018laser}%
  \BibitemOpen
  \bibfield  {author} {\bibinfo {author} {\bibfnamefont {Y.}~\bibnamefont
  {Shi}}, \bibinfo {author} {\bibfnamefont {H.}~\bibnamefont {Qin}}, \ and\
  \bibinfo {author} {\bibfnamefont {N.~J.}\ \bibnamefont {Fisch}},\ }\bibfield
  {title} {\enquote {\bibinfo {title} {Laser-plasma interactions in magnetized
  environment},}\ }\href@noop {} {\bibfield  {journal} {\bibinfo  {journal}
  {Phys. Plasmas}\ }\textbf {\bibinfo {volume} {25}},\ \bibinfo {pages}
  {055706} (\bibinfo {year} {2018})}\BibitemShut {NoStop}%
\bibitem [{\citenamefont {Edwards}\ \emph {et~al.}(2019)\citenamefont
  {Edwards}, \citenamefont {Shi}, \citenamefont {Mikhailova},\ and\
  \citenamefont {Fisch}}]{edwards_laser_2019}%
  \BibitemOpen
  \bibfield  {author} {\bibinfo {author} {\bibfnamefont {M.~R.}\ \bibnamefont
  {Edwards}}, \bibinfo {author} {\bibfnamefont {Y.}~\bibnamefont {Shi}},
  \bibinfo {author} {\bibfnamefont {J.~M.}\ \bibnamefont {Mikhailova}}, \ and\
  \bibinfo {author} {\bibfnamefont {N.~J.}\ \bibnamefont {Fisch}},\ }\bibfield
  {title} {\enquote {\bibinfo {title} {Laser {{Amplification}} in {{Strongly
  Magnetized Plasma}}},}\ }\href {\doibase 10.1103/PhysRevLett.123.025001}
  {\bibfield  {journal} {\bibinfo  {journal} {Physical Review Letters}\
  }\textbf {\bibinfo {volume} {123}},\ \bibinfo {pages} {025001} (\bibinfo
  {year} {2019})}\BibitemShut {NoStop}%
\bibitem [{\citenamefont {Muendel}\ and\ \citenamefont
  {Hagelstein}(1991)}]{muendel_four-wave_1991}%
  \BibitemOpen
  \bibfield  {author} {\bibinfo {author} {\bibfnamefont {M.~H.}\ \bibnamefont
  {Muendel}}\ and\ \bibinfo {author} {\bibfnamefont {P.~L.}\ \bibnamefont
  {Hagelstein}},\ }\bibfield  {title} {\enquote {\bibinfo {title} {Four-wave
  frequency conversion of coherent soft x rays in a plasma},}\ }\href {\doibase
  10.1103/PhysRevA.44.7573} {\bibfield  {journal} {\bibinfo  {journal}
  {Physical Review A}\ }\textbf {\bibinfo {volume} {44}},\ \bibinfo {pages}
  {7573--7579} (\bibinfo {year} {1991})}\BibitemShut {NoStop}%
\bibitem [{\citenamefont {Tang}\ \emph {et~al.}(2019)\citenamefont {Tang},
  \citenamefont {Yin}, \citenamefont {Xiao}, \citenamefont {Zhuo},
  \citenamefont {Yu}, \citenamefont {Zou},\ and\ \citenamefont
  {Shao}}]{tang2019laser}%
  \BibitemOpen
  \bibfield  {author} {\bibinfo {author} {\bibfnamefont {S.}~\bibnamefont
  {Tang}}, \bibinfo {author} {\bibfnamefont {Y.}~\bibnamefont {Yin}}, \bibinfo
  {author} {\bibfnamefont {C.}~\bibnamefont {Xiao}}, \bibinfo {author}
  {\bibfnamefont {H.}~\bibnamefont {Zhuo}}, \bibinfo {author} {\bibfnamefont
  {T.}~\bibnamefont {Yu}}, \bibinfo {author} {\bibfnamefont {D.}~\bibnamefont
  {Zou}}, \ and\ \bibinfo {author} {\bibfnamefont {F.}~\bibnamefont {Shao}},\
  }\bibfield  {title} {\enquote {\bibinfo {title} {Laser amplification by
  four-wave mixing in plasmas},}\ }in\ \href@noop {} {\emph {\bibinfo
  {booktitle} {Fifth International Symposium on Laser Interaction with
  Matter}}},\ Vol.\ \bibinfo {volume} {11046}\ (\bibinfo {organization}
  {International Society for Optics and Photonics},\ \bibinfo {year} {2019})\
  p.\ \bibinfo {pages} {110460Z}\BibitemShut {NoStop}%
\bibitem [{\citenamefont {Malkin}\ and\ \citenamefont
  {Fisch}(2020{\natexlab{a}})}]{malkin_towards_2020}%
  \BibitemOpen
  \bibfield  {author} {\bibinfo {author} {\bibfnamefont {V.~M.}\ \bibnamefont
  {Malkin}}\ and\ \bibinfo {author} {\bibfnamefont {N.~J.}\ \bibnamefont
  {Fisch}},\ }\bibfield  {title} {\enquote {\bibinfo {title} {Towards megajoule
  x-ray lasers via relativistic four-photon cascade in plasma},}\ }\href
  {\doibase 10.1103/PhysRevE.101.023211} {\bibfield  {journal} {\bibinfo
  {journal} {Physical Review E}\ }\textbf {\bibinfo {volume} {101}},\ \bibinfo
  {pages} {023211} (\bibinfo {year} {2020}{\natexlab{a}})}\BibitemShut
  {NoStop}%
\bibitem [{\citenamefont {Malkin}\ and\ \citenamefont
  {Fisch}(2020{\natexlab{b}})}]{malkin2020resonant}%
  \BibitemOpen
  \bibfield  {author} {\bibinfo {author} {\bibfnamefont {V.~M.}\ \bibnamefont
  {Malkin}}\ and\ \bibinfo {author} {\bibfnamefont {N.~J.}\ \bibnamefont
  {Fisch}},\ }\bibfield  {title} {\enquote {\bibinfo {title} {Resonant
  four-photon scattering of collinear laser pulses in plasma},}\ }\href@noop {}
  {\bibfield  {journal} {\bibinfo  {journal} {Physical Review E}\ }\textbf
  {\bibinfo {volume} {102}},\ \bibinfo {pages} {063207} (\bibinfo {year}
  {2020}{\natexlab{b}})}\BibitemShut {NoStop}%
\bibitem [{\citenamefont {Burns}\ \emph {et~al.}(2020)\citenamefont {Burns},
  \citenamefont {Vasil}, \citenamefont {Oishi}, \citenamefont {Lecoanet},\ and\
  \citenamefont {Brown}}]{burns_dedalus_2020}%
  \BibitemOpen
  \bibfield  {author} {\bibinfo {author} {\bibfnamefont {K.~J.}\ \bibnamefont
  {Burns}}, \bibinfo {author} {\bibfnamefont {G.~M.}\ \bibnamefont {Vasil}},
  \bibinfo {author} {\bibfnamefont {J.~S.}\ \bibnamefont {Oishi}}, \bibinfo
  {author} {\bibfnamefont {D.}~\bibnamefont {Lecoanet}}, \ and\ \bibinfo
  {author} {\bibfnamefont {B.~P.}\ \bibnamefont {Brown}},\ }\bibfield  {title}
  {\enquote {\bibinfo {title} {Dedalus: {{A}} flexible framework for numerical
  simulations with spectral methods},}\ }\href {\doibase
  10.1103/PhysRevResearch.2.023068} {\bibfield  {journal} {\bibinfo  {journal}
  {Physical Review Research}\ }\textbf {\bibinfo {volume} {2}},\ \bibinfo
  {pages} {023068} (\bibinfo {year} {2020})}\BibitemShut {NoStop}%
\end{thebibliography}%
\end{document}